\documentclass{aa}  
\usepackage{epsfig,graphicx,txfonts,url}
\begin{document}

\title{New technique to measure the cavity defects of Fabry-Perot interferometers}
%used in astronomical instrumentation}
\titlerunning{Cavity defects of Fabry-Perot interferometers}

\author{V. Greco \inst{1}
          \and
          A. Sordini \inst{1}
          \and
          G. Cauzzi \inst{2,3}
          \and
          K. Reardon \inst{2,3}
          \and
          F. Cavallini \inst{3}
%          \fnmsep\thanks{Just to show the usage
 %         of the elements in the author field}
          }

   \institute{Istituto Nazionale di Ottica (INO), CNR, Largo E. Fermi 3,
                  50125 Firenze, Italy\\
 \email{greco@ino.it}         
         \and
                   National Solar Observatory, 3665 Discovery Dr., Boulder CO, 80303, USA  
                     \and 
                               INAF-Osservatorio Astrofisico di Arcetri, Largo E. Fermi 5, 50125 Firenze, Italy%\\
%                   \email{gianna.cauzzi@inaf.it}     
 }

   \date{Received ; accepted }

% \abstract{}{}{}{}{} 
% 5 {} token are mandatory
 
  \abstract
   {Several astronomical instruments, for both nighttime and solar use, rely on tunable 
   Fabry-Perot interferometers (FPIs). Knowing the exact shape of the etalons' cavity
   is crucial for assessing the overall instrumental transmission profile and its possible variations 
   during the tuning process. }
   {We aim to define and test a technique to accurately measure the cavity defects of air-spaced FPIs,
   including distortions due to the spectral tuning process that are typical of astronomical observations.
   We further aim to develop a correction technique to maintain the shape of the cavity as constant as possible during the 
  spectral scan. These are necessary steps to optimize the spectral transmission profile of a two-dimensional 
   spectrograph (polarimeter) using one or more FPIs in series, and to ensure that the spectral transmission profile 
   remains constant during typical observing conditions.}
   {We devised a generalization of the techniques developed for the so-called phase-shifting 
   interferometry to the case of FPI. This measuring technique is applicable to any given FPI that can be tuned via changing the cavity spacing ($z$-axis), and can be used for any etalon regardless of the coating'
  reflectivity. The major strength of our method is the ability to fully characterize the cavity during a spectral scan, allowing for the determination of scan-dependent modifications of the plates. We have applied the measuring technique to three 50 mm diameter interferometers, with cavity gaps ranging between
      600 $\mu$m and 3 mm, coated for use in the visible range. }
  %The corrections introduced to maximize stability of the cavity  shape have been tested only on FPIs and controllers (CS100) manufactured by ICOS, and are thus valid only for this kind of instruments.
   {  The technique developed in this paper allows us to accurately and reliably measure the cavity defects of 
   air-spaced FPIs, and of their evolution during the entire spectral scan. 
   %Applying the measuring technique to several ICOS FPIs, 
   Our main, and unexpected, result is that the relative tilt between the two FPI plates varies significantly during the spectral scan, and
   can dominate the cavity defects; in particular, 
   we observe that the tilt component at the extremes of the scan is sensibly larger than that at
   the center of the scan.  Exploiting the capability of the electronic controllers to set the reference plane at any given spectral step,  we then develop a correction technique that allows the minimization of the tilt 
   during a complete spectral scan. The correction remains highly stable over long periods, well beyond the typical duration of astronomical observations.}
   {}


 \maketitle

%
%-------------------------------------------------------------------

\section{Introduction} \label{sec:Intro}

%\subsection{Context\LEt{ subsections are not allowed in the Introduction. Please remove the numbered subheadings 1.1 and 1.2.}}

Astronomical instruments based on Fabry-Perot interferometers (FPIs) have proved a flexible and efficient way to perform imaging spectroscopy and polarimetry. 
In nighttime astronomy, the first system of this kind, and one of the most successful to date, has been the Taurus Tunable Filter, which operated for a decade on the Anglo-Australian Telescope and the William Herschel Telescope \citep{1998PASA...15...44B,2002MNRAS.329..759J}. Building on this experience, a number of similar instruments have been successively developed and used  at a variety of telescopes. These currently include OSIRIS at the GTC 10-m telescope 
\citep{2013RMxAC..42...77C,2013hsa7.conf..868C};
%(Cepa et al 2013a,b);
%  Gonzalez et al 2014; Weinzirl et al 2015), are interesting followups!
the Maryland-Magellan Tunable Filter MMTF \citep{2010AJ....139..145V};
%(Veillleux 2010); 
the FP system for the Robert Stobie Spectrograph  on SALT \citep{2008AJ....135.1825R,2016SPIE.9908E..6NW}.
%(Ragwala 2008; Williams et al 2016).
New instruments are also being planned, such as the Brazilian Tunable Filter Imager (BTFI2) for the SOAR telescope \citep{2018SPIE10702E..57Q},
%(Quint et al. 2018), 
or a system for detection of molecular oxygen in exoplanets' atmosphere operating on forthcoming ELTs  \citep{2018SPIE10702E..6NB}.
%(Ben Ami 2018). 

Among the most-used instruments in solar physics are 
 the dual FPI CRisp Interferometric SpectroPolarimeter \citep[CRISP,][]{2008ApJ...689L..69S}
 %Scharmer et  al 2008) 
 and CHROMIS (Scharmer 2019, in prep.), both at the Swedish Solar Tower (SST); 
 the Triple Etalon SOlar Spectrometer  \citep[TESOS,][]{1998A&A...340..569K,2002SoPh..211...17T}
% Kentischer et al 1998; Tritschler et al 2002) 
    at the German Vacuum Tower Telescope (VTT); the dual FPI Italian Interferometric BIdimensional 
    Spectrometer installed at the Dunn Solar Telescope of the National Solar Observatory \citep[IBIS,][]{2006SoPh..236..415C};
%    Cavallini 2006); 
the Imaging Magnetograph eXperiment (IMaX)  flown on the Sunrise I and II balloons \citep{2011SoPh..268...57M};
%Martinez Pillet et al. 2011); 
and the Gregor Fabry Perot Interferometer currently operating at the German GREGOR 
    1.5 m telescope in Tenerife \citep[GFPI,][]{2012AN....333..880P}.
% Puschmann et al 2012).  
The Visible Tunable Filter is currently under construction, planned for installation at the upcoming  4-m
%\LEt{ change made for consistency with 10-m above.} 
aperture solar 
 telescope DKIST \citep{2016SPIE.9908E..4NS,2018SPIE10700E..0VW}.
 %Schmidt et al 2016; Warner et al 2018). 
    
 Most of these FPIs are air-spaced, and spectrally tunable by rapidly changing 
 the cavity spacing via piezo-electric actuators. A peculiar characteristic of the FPIs used for solar instruments is  their very high spectral resolution ($R >$100,000), which allows precise spectropolarimetry of narrow spectral lines. This requirement
  translates into FPI gaps on the order of one to few millimeters, in contrast with cavities of few tens of microns, for the majority of nighttime instruments. Instruments using these FPI rapidly and repeatedly change the gap of the FPI over a wide range to select different transmission wavelengths and thus perform a spectral scan. 

The overall transmission profile of these instruments, and its possible variation with wavelength and across the 
    field of view (FOV) are important parameters in order to properly interpret the observed spectral and polarimetric profiles. In particular, the widely used technique of spectral inversions as a way to derive 
    solar atmospheric parameters from the shape of the observed spectral lines \citep{1992ApJ...398..375R,2017SSRv..210..109D}
 %   (Ruiz Cobo \& del Toro Iniesta 1992; 
  %  de la Cruz Rodriguez \& van Noort 2017) 
  requires an exact knowledge of the shape of the instrumental 
    transmission profile, as details like asymmetries in the observed Stokes profiles might reveal 
    gradients in physical quantities such as velocity, or magnetic field intensity and direction. As an example, an earlier work by 
    \citet{2008A&A...481..897R}
    %Reardon \& Cavallini (2008) 
    assesses the influence of the cavity errors of the FPIs employed in 
    IBIS on the overall instrumental profile. By using simulated spectral profiles of a photospheric line, they show how using the ``nominal'', ideal instrumental 
    profile rather than the true one, could introduce uncertainties in the convolved line profiles comparable 
    with the expected signal, for example for the case of weak polarization signals (at the level of few $\times 10^{-3}$ 
    the continuum intensity). For instruments using the telecentric mount, defect-induced wavelength shifts  across the FOV are also 
    an important factor to consider, as they compound other issues such as the need to obtain the highest possible spatial resolution via image reconstruction techniques \citep[e.g.,][]{2015A&A...573A..40D}.
    %(e.g.  de la Cruz Rodriguez et al. 2015).
 %CRISPRED: between acquisitions, solar features are moved and distorted on the FOV, so there is not a unique wavelength shift (or pixel) that can be associated with that feature as it may have sampled a range of pixels in the different exposures and wavelengths. 

%\subsection{Cavity errors\LEt{ please remove this subheading.}}

The cavity errors of a FPI can be decomposed in large and small scale components. The small-scale errors, which have
     negligible size with respect to the aperture of the FPI plates, are essentially due to the 
     microroughness of the glass substrate and/or the coating. The large-scale errors, 
     meaning those with sizes comparable to the size of the plates, can be due to a variety of factors. These include manufacturing 
     errors; tension on the plates due to the coating;
       pre-load stresses introduced in the assembly of the FPIs; pressure exerted by the 
      actuators; gravity; and, most important, a tilt of the plates, that is, deviations from the parallelism of the 
two planes best fitting the plates' shape. These large-scale errors are best described with a 
      series of Zernike polynomials \citep{1992aooe...11....2W}.
      %Wyant \& Creath,1992).

While most of the cavity errors are permanently determined during the fabrication phase, 
      the relative tilt of the two plates composing the cavity can be minimized and in principle eliminated during operations,
       if an accurate defects map is available, and the system includes a set of actuators that can act 
       differentially on the plates. The latter is the case for most of the FPIs used in operational astronomical instruments;
 to date, most of the tunable FPIs used in astronomy are capacity-controlled devices produced by 
IC Optical Systems Ltd. (ICOS, http://www.icopticalsystems.com/).

Various authors have described ways to measure and minimize the relative tilt of FPI' plates, 
  most often as part of the (daily) instrumental calibration at the telescope. 
\citet{1998PASP..110.1059J} and \citet{2004SoPh..220...21M}
%    Jones \& Bland-Hawthorn (1998) and Mickey (2004) 
proposed  using differential measurements across separate quadrants of the etalons' plates, while  
 \citet{2010AJ....139..145V}     
 % Veilleux et al (2010) 
 used a scan of both axes of movement to identify the position where the spectral profile of a  
 bright emission line from an arc lamp  is the narrowest.
 %and most symmetric. GC: Tolto, ma a me sembra che nel loro articolo loro facciano cenno anche alla asimmetria
 The same technique was also adopted for the SALT RSS FP system \citep{2008AJ....135.1825R}
%       (Ragwala et al 2008) 
but added significantly to the downtime of the instrument. In a follow up work, \citet{2016SPIE.9908E..6NW}
%Williams et al (2016) 
described how the instrumental setup could be accelerated by exploiting the correlation of the tilt parameters with the position of the center of a calibration interference ring. A similar idea was presented by \citet{2005PASP..117.1435D}, who observe how the tip-tilt parameters characterizing the cavity are linearly dependent on the voltage defining either axis of
movement; a single measure of the cavity shape (defined in terms of Zernike polynomial, see below) would then be sufficient to
derive the optimal settings for minimizing the tilt.
     For the case of IBIS, \citet{2006SoPh..236..415C}
     %Cavallini (2006) 
     describes the procedure adopted to minimize the parallelism error of each FPI separately using the calibration channel of the instrument, which includes a laser source and a diffuser, coupled with a set of lenses that can image the cavities of the two FPIs (one at a time) onto the science camera. 
      
 Although rapid and efficient, most of these procedures
 are  based on approximate estimates of the best orientation of the plates, and are often amenable 
 to subjective interpretation. 
For example,  \citet{2008A&A...481..897R}
%Reardon \& Cavallini (2008) 
perform a full analysis of the cavity errors of IBIS and showed that
after the manual tilt minimization procedure they could still measure a significant residual tilt, with peak-to-valley ($PV$) values in the range 
\begin{equation} 
\centering
             1.24~$nm$ ~\le PV \le 3.17~$nm$      
\label{eq:IBISresidualtilt} %Eq 1                                             
.\end{equation}
\noindent

For a single FPI in a collimated mount, such a residual tilt produces a broadening of the transmission profile approximately expressed by the relation 
\begin{equation} 
\frac{\ \ \delta FWHM \ \ }{ FWHM}= \sqrt{ \ \  1+3R \ \Bigg [ \frac{\pi \  \Delta t_M}{\ \ \lambda \ (1-R) \ \ } \Bigg ]^2 \ \  } - \ 1
\label{eq:DeltaFWHM} %Eq 2
,\end{equation}

\noindent
where FWHM is the full width at half maximum of the transmission profile of an ideal interferometer (no cavity defects); 
$\Delta t_M$ is the $PV$ value of the tilt; R the reflectivity of the plates, and $\lambda$ the wavelength. If in Eq. \ref{eq:DeltaFWHM}
 we assume R = 0.95 (typical of solar instruments) and $\lambda$=550 nm, we obtain a broadening of 3\% for $\Delta t_M$ =1.24 nm, and of 17\% for 
$\Delta t_M$ =3.17 nm.  In a telecentric mount, the same values of residual tilt would introduce shifts of 
the spectral transmission profile on the order of 200-500 ms$^{-1}$ (at visible wavelengths) over the whole field of view. These are already significant
degradations of the ideal instrumental profile but, even 
more importantly, they will compound in a non-trivial manner, in the common case 
of multiple interferometers operating in series, for example decreasing the overall transmission or changing the shape of the combined profile
\citep[e.g.,][]{2016SPIE.9908E..6NW,2018SPIE10702E..6NB}.
%(e.g. Williams et al 2016; Ben-Ami et al 2018). 
 
A more objective characterization of the cavity errors of FPI, and a more accurate way to control the plate parallelism, 
appears thus necessary. In the present paper, we have followed the approach of \citet{2005PASP..117.1435D}
%Denker \& Tritschler (2005) 
and \citet{2008A&A...481..897R},
%Reardon \& Cavallini (2008), 
to fully describe the cavity shape of a Fabry Perot Interferometer in terms of Zernike polynomials. Extending their analysis, however, we explicitly consider the 
possibility that the cavity errors might change during the full spectral scan. This is motivated by two distinct factors. First, 
some deformation of the cavity can be expected when it is pushed far from the central gap spacing,  because
of the relatively large excursions of the piezoelectrics and the FPI's plates during the tuning process, while confined in an essentially rigid, overall hardware structure. The extremes of the spectral scan can often be approached when exploring a variety of astronomical  problems, so this effect is worth investigating. Second, even though the
overall alignment of the FPI's plates is supposed to be maintained by the balancing action of the capacitance bridge used to drive the piezo-electric actuators (see e.g., Sect. \ref{sec:correction}), our measurements on several ICOS FPIs, described
below, show that this might not be the case.  
%
%{\bf despite the balancing action of the capacitance bridge used to drive the piezo-electric actuators (see e.g. Sect. \ref{sec:correction}).}
%in  perhaps due to varying responses of the actuators, or stresses induced by those same actuators. 
%
To our knowledge, this is the first time that  possible changes of the cavity errors during the spectral scan are  explicitly taken into account.

In Sect. \ref{sec:Cavity} we present a novel procedure,
developed and tested at the Optical Measurements and Testing Laboratory of the Italian National Center for 
Research (CNR - INO), to measure the cavity defects of any given FPI of up to 
150 mm diameter. In Sect. \ref{sec:casestudy} we describe in detail the application of this technique to an existing 50 mm 
diameter FPI (model ET50, fabricated by ICOS) and the ensuing results. We applied the same technique on two other ICOS ET50, obtaining consistent results; we provide these in Appendix A. %for sake of clarity.  
As shown in Sect. \ref{sec:casestudy}, the plates' tilt, and hence the transmission profile, 
can indeed vary significantly within the spectral scan; 
in Sect. \ref{sec:correction} we further define and test a procedure 
to minimize and stabilize the tilt during a full spectral scan, discussing the resulting effects on the instrumental transmission profile.        
Finally, we present our discussion and conclusions in Sect. \ref{sec:discussion}.
  
%--------------------------------------------------------------------
\section{Characterizing the FPI cavity} \label{sec:Cavity}

\subsection{The measuring technique: analytical formulation} \label{sec:definitions}

%% Figure 1
\begin{figure} 
\centering
  
\includegraphics[width=8cm]{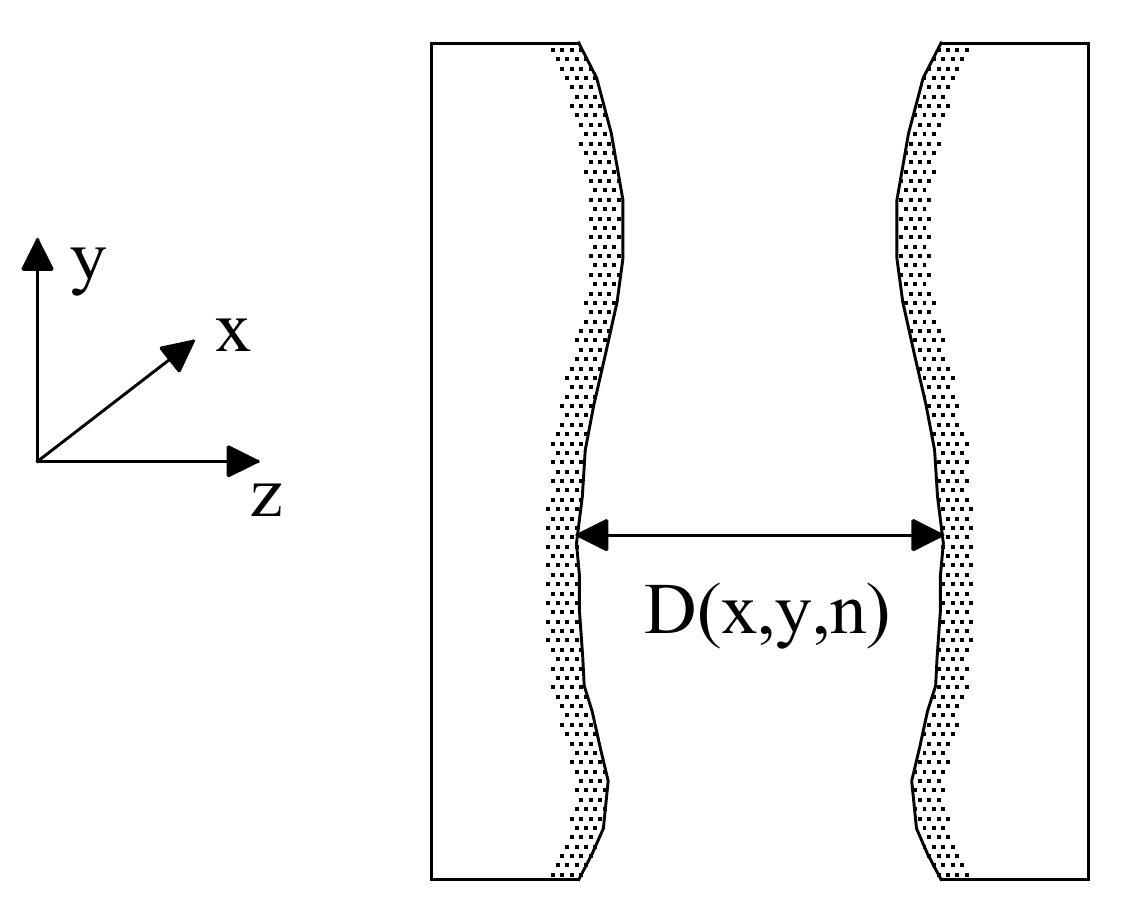}
   \caption{Optical cavity of a non-ideal Fabry-Perot interferometer. $n$ is the step position within the spectral scan ($n \in [-2047,2048]$ for a typical ICOS controller. See also Eq. \ref{eq:DefD} below).}   
   \label{fig:figcavity}
 \end{figure}

Figure \ref{fig:figcavity} shows a schematic non-ideal, air-spaced Fabry-Perot interferometer, 
where the two plates that combine to form the actual cavity are not perfect parallel planes. 
The defects are enhanced for clarity. In all of our analyses, we have assumed that the index of refraction of the air
within the etalon remains constant; this
holds for our laboratory measurements but also for typical instruments (e.g., IBIS) where the interferometers are pressure- and temperature-controlled (to within $<$ 0.1$^\circ$ C in the case of IBIS).

%this is typically the case in operating instruments (e.g. IBIS), where the interferometers are 
%encased in a sealed, temperature controlled container. }

We call the cavity spacing  D$_0$, defined as the average of the distance between the single plate points, 
over the whole surface. A system of actuators (three in the case of ICOS ET50, see Fig. \ref{fig:actuators} below) 
allows us to change the plates' separation along the longitudinal axis ($z$ in Fig.  \ref{fig:figcavity}); this is how spectral tuning is 
achieved in most current instruments. Using a proximity sensor system, the actuators' controller shifts one plate with respect to the other in constant steps, of size $\Delta$, within the range 
\big [ D$_0 - \big (2^{(N-1)} -1 \big ) \cdot \Delta$, D$_0 + \big (2^{(N-1)} \big )  \cdot \Delta$ \big ], 
where $N$ is the digital resolution of 
the controller. For the case of high spectral resolution ICOS interferometers, typical values are: 
D$_0$ = one to few mm; $N$ = 12; and $\Delta$ = 0.5 nm.

Using the reference system $ (x, y, z)$ depicted in Fig. \ref{fig:figcavity}, with the $z$ axis coinciding with the optical axis of the interferometer, the distance $D(x,y,n)$ between any two points of coordinates $(x,y)$ on the plates  is given by:
\begin{equation} 
D(x,y,n) \ = \ D_0 \ + \ n \cdot \Delta \ + \  F(x,y,n)
\label{eq:DefD} % Eq 3
,\end{equation}
where $n \in \big [-\big(2^{(N-1)}-1\big), \ 2^{(N-1)} \  \big ]$ is any given step of the scan, and the map $F(x,y,n)$ includes all the cavity errors. Assuming that  $F(x,y,n)$ is a slowly varying function of $n$, 
%i.e. that the cavity errors are mostly determined during the fabrication phase,
% and do not change during a spectral scan, 
we can write:

\begin{equation} 
F(x,y,n) \ \simeq \ F_0(x,y) \ +  \  F_1(x,y) \cdot n \ +  \  F_2(x,y) \cdot n^2
\label{eq:SviF} %Eq 4
,\end{equation}

\noindent
where the term $F_0 (x,y)$ includes all the static cavity errors apart from those introduced by the moving actuators, encoded in $F_1 (x,y)$ and $F_2 (x,y)$.

Substituting Eq. \ref{eq:SviF} in Eq. \ref{eq:DefD} we obtain:

\begin{equation} 
D(x,y,n) \ = \ D_0 \ + \ \big[ \Delta + F_1 (x,y) \big] \cdot n \ + \   F_2(x,y) \cdot n^2 \ + \ F_0 (x,y)
\label{eq:DefDapprox} %Eq 5
.\end{equation}

To first order in $n$, for each cavity pixel $(x,y)$ the spectral scan thus results in a different step size, $ \Delta_1 $ given by:

\begin{equation} 
\Delta_1 \ = \ \Delta \ + \ F_1(x,y)
\label{eq:Delta1} %Eq 6
.\end{equation}

Further, since the map of cavity defects $M(x,y,n)$ is defined (apart from an additive constant) by:

\begin{equation} 
M(x,y,n) \ = \ D(x,y,n) \ - \ D(0,0,n)
\label{eq:DefM} %Eq 7
\end{equation}

\noindent
using Eq. \ref{eq:DefDapprox} , we obtain:

\begin{multline}
M(x,y,n)  \simeq \  \big[ F_1 (x,y) - F_1(0,0) \big] \cdot n \ + \   
                             \big[ F_2 (x,y) - F_2(0,0) \big] \cdot n^2 \\
                             + \ \big[ F_0 (x,y) - F_0(0,0) \big] 
\label{eq:DefMapprox}. %Eq 8
\end{multline}

By illuminating the FPI with a monochromatic, collimated beam along the $z$ axis, and assuming negligible absorption in the cavity, the transmitted $I_T (x,y,n)$ and reflected $I_R (x,y,n)$ intensities are given by the well known Airy relations 
\citep[e.g.,][]{1989fpih.book.....V}:
%(e.g. Vaughn 1989):

\begin{gather}
I_T(x,y,n) \ = \ \frac {I_0(x,y)} 
                    {\ \ 1 + \dfrac {2 R} {\ \ (1-R)^2 \ \ }
                      \bigg[ 
                      1 - \cos \bigg( \dfrac{4 \pi}{\lambda} D(x,y,n) \bigg)
                      \bigg] \ \ 
\label{eq:IT}                       %Eq 9
                    } \\
                   I_R(x,y,n) = I_0(x,y) - I_T(x,y,n)
\label{eq:IR}             % Eq 10      
\end{gather}

\noindent
where $I_0 (x,y)$ is the incident intensity, $\lambda$ is the wavelength, and $R$ the reflectivity of the plates (a slowly varying function of  $\lambda$).

If we measure the intensity of the transmitted and reflected beams with a 2-D detector (e.g., a CCD), as  $n$  is stepped
through increasing values every pixel $(x,y)$ will experience a succession of interference orders with modulation:

\begin{multline} 
K_T = \frac {(I_T)_{max} - (I_T)_{min}}{(I_T)_{max} + (I_T)_{min}} = \frac {2R}{1+R} \\
K_R = \frac {(I_R)_{max} - (I_R)_{min}}{(I_R)_{max} + (I_R)_{min}} = 1  \ \  \ (I_R)_{min}=0
\label{eq:DefK} %Eq 11
\end{multline}

\noindent 
%Using the previous equations, 
$I_T$  and $I_R$ can be written as

\begin{multline}
I_T(x,y,n) \ =  C(x,y) \ + \\
                    \frac {A(x,y)} 
                    {\ \ 1 + B(x,y)  \cdot 
                      \Big \{ 
                      1 - \cos \Big[  
                      \alpha(x,y) \cdot n + \beta(x,y) \cdot n^2 + \gamma(x,y)
                      \Big]
                      \Big \} 
                     }
                     \label{eq:ITfit} %Eq 12
\end{multline}            

\begin{multline}
I_R(x,y,n) \ =  C(x,y) \ + \ A(x,y) \cdot \Bigg \{ \ \ 1 \ - \\
                    \frac {1} 
                    {\ \ 1 + B(x,y) \cdot
                      \Big \{ 
                      1 - \cos \Big[  
                      \alpha(x,y) \cdot n + \beta(x,y) \cdot n^2 + \gamma(x,y)
                      \Big]
                      \Big \} 
                     } 
                     \Bigg \}
                     \label{eq:IRfit} %Eq 13
\end{multline}   

\noindent
where:
\begin{equation}
\begin{split}
\alpha(x,y) \ &= \ \frac{4 \pi} {\lambda} \cdot \big[  \Delta + F_1(x,y) \big]  \\
\beta(x,y) \ &= \ \frac{4 \pi} {\lambda}  \cdot  F_2(x,y)  \\
\gamma(x,y) \ &= \ \frac{4 \pi} {\lambda} \cdot \big[  D_0 + F_0(x,y) \big] 
\label{eq:albega} %Eq 14
\end{split} 
.\end{equation}   

In Eqs. \ref{eq:ITfit} and  \ref{eq:IRfit}, the matrix $A(x,y)$ describes possible spatial variations of both the incident beam's intensity and the gain by each CCD pixel; matrices $B(x,y)=\tfrac{2R}{(1-R)^2}$  and $C(x,y)$ describe spatial variations of the plates' reflectivity and pixels' offset, respectively. 

Let us assume we acquired an interferogram $I_R (x,y,n)$ of the reflected beam (alternatively, of  the transmitted beam) for every step $n$ of the spectral scan. By fitting the intensity measured in each pixel $(x,y)$ as a function of $n$ using Eq. \ref{eq:IRfit}, we can determine the six matrices: $A(x,y)$, $B(x,y)$, $C(x,y)$, $\alpha(x,y)$, $\beta(x,y)$ and $\gamma(x,y)$, and hence the way in which the map of cavity errors evolves during the spectral tuning. In fact, assuming that the $PV$ value of the cavity errors is $ < \lambda/2$, substituting  Eq. \ref{eq:albega}  in Eq. \ref{eq:DefMapprox} we obtain: 

\begin{multline}
M(x,y,n) = H \cdot \frac {\lambda}{4 \pi} \cdot \Bigg \{  \\
               \Big[ \alpha(x,y) -\alpha(0,0) \Big] \cdot n +  \Big[ \beta(x,y) -\beta(0,0) \Big] \cdot n^2 
                +  \Big[ \gamma(x,y) -\gamma(0,0) \Big] 
                \Bigg \}
\label{eq:difetto} %Eq 15
\end{multline}

\noindent
with H an ``inversion'' factor that can be either $[+ 1,-1]$. In the first case (H=$+1$), $M(x,y,n)$ describes the map of the cavity defect (apart from an additive value), while in the second (H=$-1$) $M(x,y,n)$ describes the shape of the surface of one of the plates, assuming the other is a perfect plane. This is equivalent to saying that the knowledge of $I_R$ or $I_T$ is not sufficient to separately determine the shape of the two surfaces composing the cavity. 

The technique to measure cavity defects that we describe here below is based on Eqs.  \ref{eq:ITfit}-\ref{eq:difetto}, and can be 
considered as a generalization of the techniques developed within the so-called phase-shifting interferometry 
\citep{schreiber2007phase}
%Schreiber & Bruning 2007
to the case of FPIs. The phase-shifting technique has been developed within the 
framework of optical testing, and is usually employed to characterize cavities that: $i$) have a very low reflectivity $(R \simeq 0.04)$, 
so that the interference orders are essentially sinusoidal; and $ii$) are supposed to have constant shape throughout the spectral 
scan. Only recently some modifications have been proposed to the classical technique, that allow it to be used in cases when the tilt of 
the cavity varies within the spectral scan \citep{2014ApOpt..53.4628D}.
%(Deck 2014). 
Our own technique expands on these efforts, in the sense that it can be used 
for any value of the reflectivity R, and can fully characterize the cavity during a spectral scan, hence not only measuring the tilt 
variations, but also scan-dependent modifications of the plates (Sections \ref{sec:casestudy} and \ref{sec:correction} below). These features make 
our technique extremely useful for the characterization of astronomical Fabry-Perot, as R can be rather large $(R>0.8)$, and the 
residual tilts be of large relevance for accurate spectroscopy (see the discussion in the Introduction of this paper). 

Finally, it is trivial to prove that 
our technique is a generalization of that used in \citet{2005PASP..117.1435D} and \citet{2008A&A...481..897R}. These authors
derive the value of the step $\Delta$ by measuring the average number of steps between two consecutive maxima of the $I_T$
curve, and imposing that the distance between the maxima is $\lambda/2$, with $\lambda$ the wavelength of the laser. For every pixel $(x,y)$ of the final image they then identify (for a given interference order) the position $n_m(x,y)$ of the local $I_T$ maximum;
to this end \citet{2005PASP..117.1435D} use a gaussian fit of the intensity curve, while \citet{2008A&A...481..897R}
employ a center of gravity method. 

The map of cavity defects $M_0[x,y,n_m(0,0)]$ corresponding to the mid-step of the scan, $n_m(0,0)$, is then computed
using:

\begin{equation} 
M_0\Big[x,y,n_m(0,0)\Big] \ = \ H \cdot \Delta \cdot \Big[n_m(0,0) - n_m(x,y) \Big] 
\label{eq:DefM0}
\end{equation}

\noindent
with $H$ defined as in Eq. \ref{eq:difetto}. By comparing with Eq.  \ref{eq:ITfit}, we see that the matrix $ n_m(x,y)$ is
defined by:

\begin{equation} 
\alpha(x,y) \cdot n_m(x,y) + \beta(x,y) \cdot n_m^2(x,y) + \gamma(x,y) = 2 k \pi 
\label{eq:nmax}
\end{equation}

with $k=0, \pm 1, \pm2,...$ depending on the interference order. If we derive 
$\gamma(x,y) $ from Eq.  \ref{eq:nmax}, and substitute it in Eq. \ref{eq:difetto}, using Eq. \ref{eq:albega} we obtain:
\begin{multline}
M\Big[x,y,n_m(0,0)\Big] \ = \ H \cdot \Bigg\{  
 \Big[\Delta + F_1(x,y) \Big] \cdot \Big[n_m(0,0) - n_m(x,y)\Big] + \\ 
F_2(x,y) \cdot \Big[n_m^2(0,0) - n_m^2(x,y)\Big]
\Bigg\}
\label{eq:difetto2}.
\end{multline}
For $F_1(x,y) = F_2(x,y) \equiv 0$  Eq. \ref{eq:difetto2} is equal to Eq. \ref{eq:DefM0}, thus demonstrating that our technique is
equivalent to the earlier ones if we neglect the possibility of cavity distortions during the scan.

\subsection{Measuring technique: experimental setup} \label{sec:Zygo}

\begin{figure*} [ht]
\centering
%  \vspace{8cm}
\includegraphics[width=16cm]{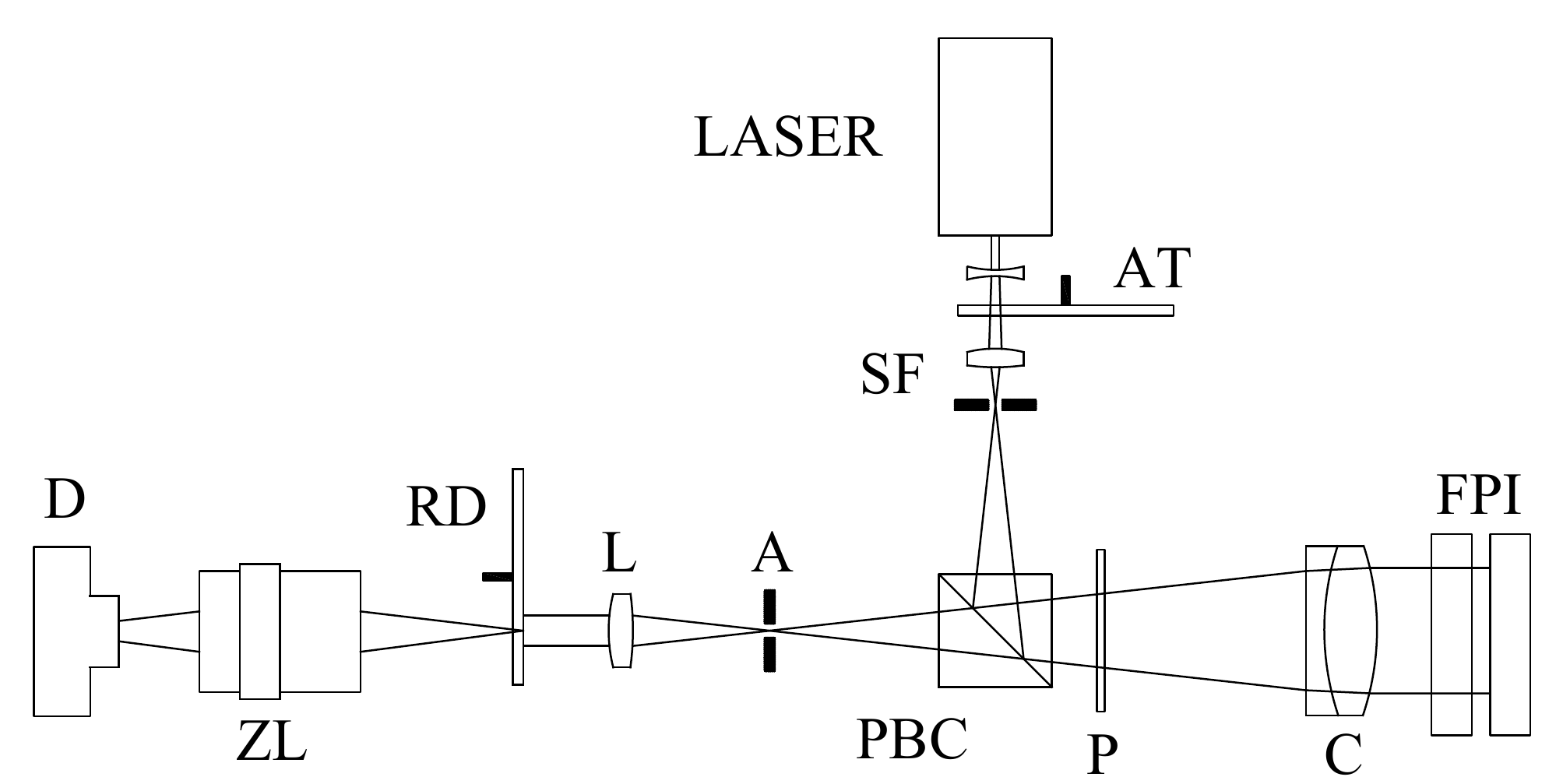}
   \caption{Optical setup adopted to measure the reflected intensity from a Fabry-Perot interferometer FPI. 
   AT = Attenuator; SF = Spatial Filter; PBC = Polarizing Beam Splitter Cube; P = $\lambda /4$ Plate; C = Collimator; 
   FPI = Fabry-Perot interferometer; A = Aperture (diaphragm); L = Imaging Lens; RD = Rotating Diffuser Disk; 
   ZL = Zoom Lens; D = Detector.}    
   \label{fig:figZygo}
 \end{figure*}
 
In Fig. \ref{fig:figZygo} we show the optical setup adopted to measure the intensity reflected by a FPI. The spatial filter (SF) 
produces a point-like monochromatic source, in the focus of the collimator (C). The laser beam  is linearly 
polarized, and oriented so as to be completely reflected by the polarizing beam splitter cube (PBC). The beam is then directed through the $\lambda /4$ plate (P), oriented at 45$^{\circ}$ with respect to the direction of polarization, 
thus emerging circularly polarized. The beam reflected by the FPI, after passing through the collimator and the quarter-wave plate is again linearly polarized, but at 90$^{\circ}$ with respect to the original direction, and is hence transmitted by the PBC, 
and focused on the diaphragm A. The diaphragm is necessary to filter out all the spurious fringes produced 
by all the surfaces within the optical path apart from the internal surfaces of the FPI. The lens 
L remaps the optical cavity of the FPI onto a rotating diffuser disk, which is needed to remove the speckle field 
introduced by the laser. Finally, the zoom lens (ZL) images the cavity onto a 512$\times$512 camera (D).

The FPI transmitted intensity could be measured with an analogous, simpler optical setup, but the measure of the 
reflected intensity is more versatile. Indeed, while for high values of reflectivity (R $>$ 0.9) the two setups would be equivalent, 
for low values (R $\sim$ 0.04) the measure of the reflected intensity is more advantageous since the modulation of the 
transmitted beam is about  8\%, vs. the 100\% of the reflected beam (Eq. \ref{eq:DefK}). Examples of low reflectivity 
cases include an uncoated cavity, useful to characterize the fabrication errors separately from the coating ones (R $\sim$ 4\%, 
typical of the air-glass system), or a coated cavity optimized for wavelengths far from that of the laser used for
 the measures (e.g., infrared-optimized coatings). 

\section{Experimental data: the case study of an ICOS ET50} \label{sec:casestudy}

As a case study, we applied the technique described in Sect. 2
%\LEt{ please check.} 
to the measure of 
the cavity defects of a servo-stabilized interferometer system, composed of a ICOS FPI ET50 FS 
and its CS100 controller. 
%Both are manufactured by the IC Optical Systems Ltd. (http://www.icopticalsystems.com/); 
%to date, most of the capacity-controlled, tunable FPIs used in astronomy are produced by 
%ICOS, with diameters between 50 and 150 mm. 
The FPI ET50 and CS100 were previously used as
the main components of the Interferometric Panoramic Monochromator installed at the THEMIS telescope 
\citep[IPM,][]{1998A&AS..128..589C,1999A&A...344L..29B}.
%(IPM, Cavallini, 1998;  Berrilli et al. 1999). 
The relevant characteristics of the system are given in Table \ref{tab:et50properties}.

\begin{table}
\caption{Servo-Stabilized Interferometer System characteristics}
 \label{tab:et50properties}
\centering
\begin{tabular}{l l }
\hline 
\multicolumn{2}{c}{Fabry-Perot Interferometer}\\
\hline
Manufacturer &IC Optical Systems\\
Type & ET50 FS \\
Aperture & 50 mm \\
Material & Fused Silica\\
& (minimal inhomogeneity)\\
Plate Wedge & 20 arcmin \\
Cavity Spacing &3.000 mm $\pm \sim$0.005 mm\\
Coating &Multilayer broadband\\
Wavelength range & 400 \ nm $-$ 700 \ nm\\
Estimated cavity errors& $\lambda$ / 100 (PV $@$ 632.8 nm, after\\
&coating, over central 35 mm)\\
Nominal Reflectance & 95\% $\pm$ 3\% \\
Estimated Step $(\Delta)$ & $\sim$0.46 nm\\
Tilt Control Estimated precision (X, Y)  & $\sim$1.4 $\cdot 10^{-3}$ arcsec\\
%$(\Delta_\theta)$ & $\sim$1.4 $\cdot 10^{-3}$ arcsec\\
%Estimated Tilt Y Step $(\Delta_\theta)$ & $\sim$1.4 $\cdot 10^{-3}$ arcsec\\
\hline
\multicolumn{2}{c}{Controller}\\
\hline
Manufacturer &IC Optical Systems\\
Type & CS100 \\
Digital Resolution (N) & 12 bits / 4096 steps\\
\hline
\end{tabular}
\end{table}

The measurements were performed at the ``Laboratorio di Misure e Collaudi Ottici'' of the Italian Istituto Nazionale di Ottica 
(INO), in a clean room with constant temperature and humidity (T=20$\pm$0.1 C, RH=45\% $\pm$ 5\%). This ensured that the
index of refraction of the air between the FPI plates remained constant throughout the measures.

\begin{figure} 
\centering
\includegraphics[width=8cm]{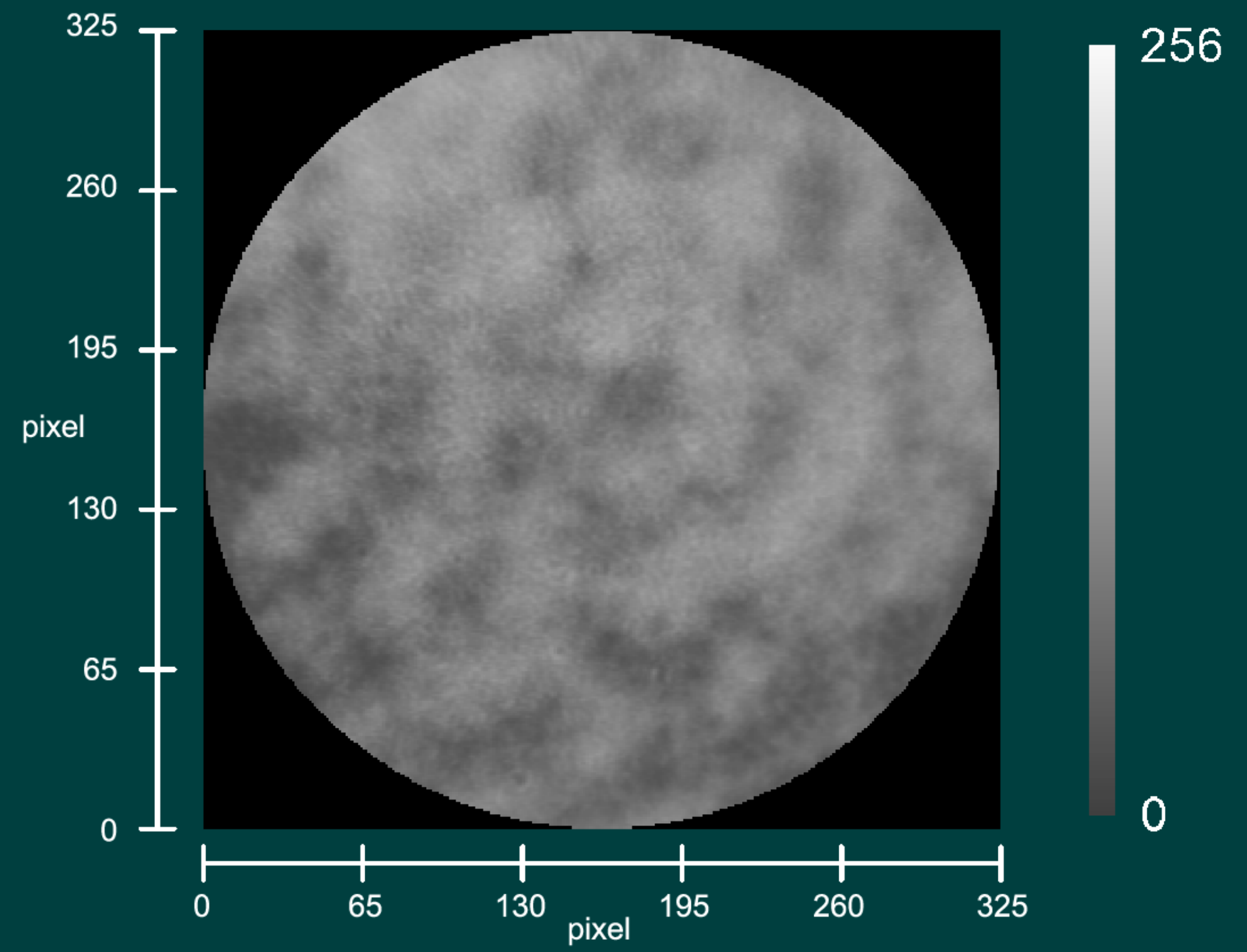}
   \caption{Typical interferogram of the ET50 measured in reflection. $\lambda$ =632.8 nm,  $n$=0}    
   \label{fig:interferogram}
 \end{figure}

The optical setup of Fig. \ref{fig:figZygo} was realized using a phase-shifting interferometer 
GPI-XP of Zygo-Ametek (www.zygo.com), that outputs  a collimated HeNe ($\lambda$=632.8 nm) laser beam 
of 100 mm diameter (a 150 mm beam expander is also available, and can be used to extend the measures to larger format FPI). 
The ET50 is positioned in front of the GPI-XP  so that the $z$-axis of the cavity 
is horizontal (perpendicular to gravity), and aligned so that its optical axis coincides with that of the collimator. 
The capability of the GPI-XP to introduce a phase-shift on the beam was disabled; rather, the GPI-XP was used 
simply as a collimator and to acquire (with 8 bit digitization) the light reflected from the 
ET50. All the elements of the setup depicted in Fig. \ref{fig:figZygo}, apart from the FPI, are contained in the GPI-XP case. Fig. \ref{fig:interferogram} shows a typical interferogram observed at the central position ($n$=0) of the spectral scan. 
The acquisition software selects a central circular section of the image, with 325 pixel diameter; 
this corresponds to the central part of the pupil of the ET50, with a diameter of 35 mm (108 $\mu$m / pixel). Comparing 
Fig. \ref{fig:interferogram}  with the equivalent images in \citet{2008A&A...481..897R}
%Reardon \& Cavallini (2008) 
we observe a diminished ``granularity'' in the cavity
errors in our case; this could be due either to the reduced dimensions and dynamic range of our detector, or to an 
intrinsic difference in the property of the coating. The corresponding maps
derived for the case of the two additional NSO FPIs show variations on smaller spatial scales (see figures in Appendix); as a result, we
favor the second hypothesis.

Using the resident software of the GPI-XP (MetroScript Programming Language) it is possible to simultaneously manage 
remotely both the GPI-XP and the CS100 controlling the ET50. To measure the cavity defects, we developed a script that 
incrementally changes the distance between the FPIs' plates by one step at a time ($\Delta$, see Table 1), and then acquires 
the corresponding interferogram with the GPI-XP. The complete spectral scan (4096 steps) takes approximately 24 minutes. 
The interferogram's acquisition starts about 360 ms after each incremental step is applied; this is much longer than the 
damping time of the cavity ($\le$ 20 ms) that we measured in the laboratory.

\subsection{Fringe fitting} \label{sec:fringefitting}

\begin{figure} 
\centering
%  \vspace{8cm}
\includegraphics[width=9cm]{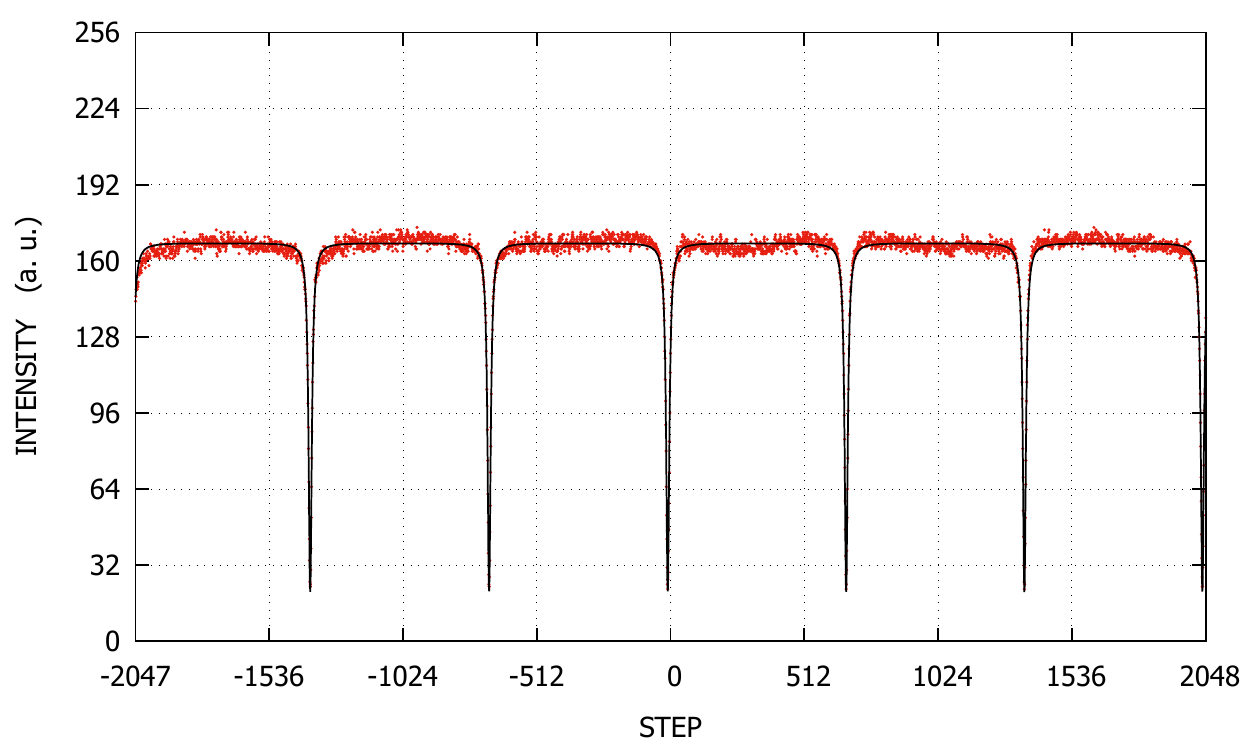}
   \caption{Intensity recorded in the central pixel of the images, during a complete spectral scan of 4096 steps. 
   Red points: recorded data points; black line: fit using Eq. \ref{eq:IRfit} }    
   \label{fig:orders}
 \end{figure}

\begin{figure} 
\centering
%\vspace{8cm}
\includegraphics[width=9cm]{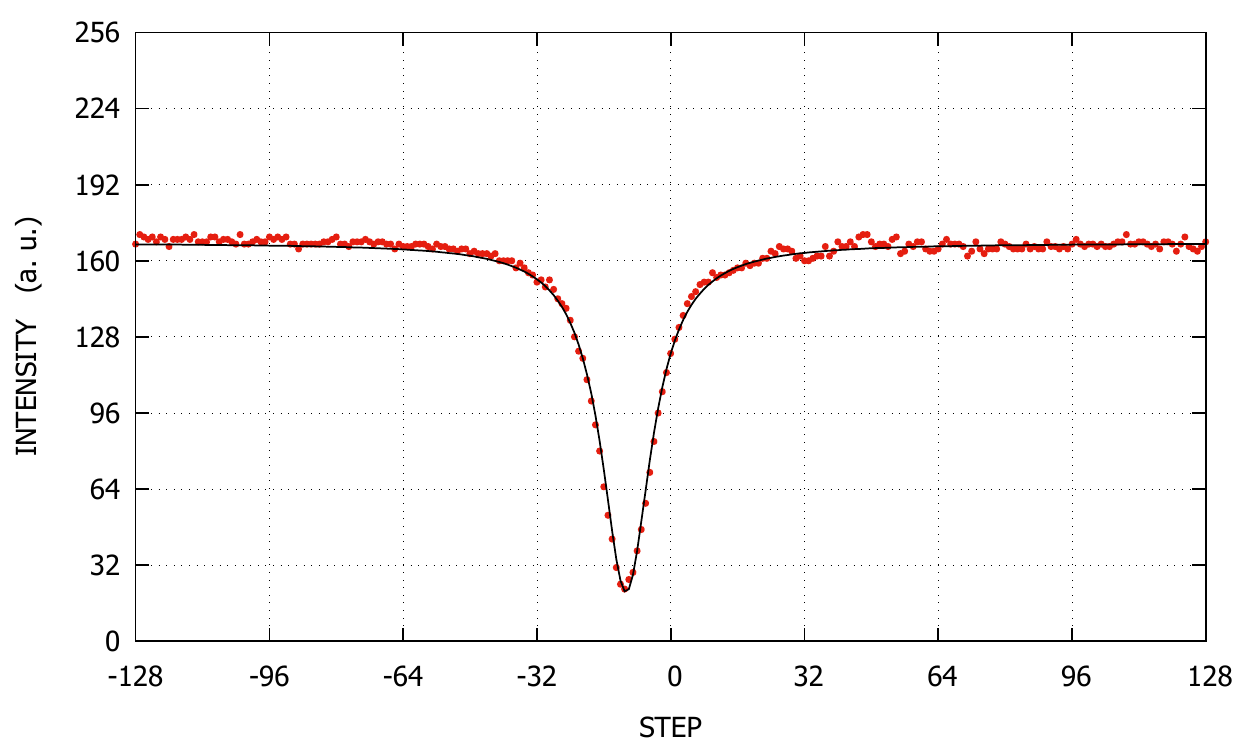}
   \caption{Detail of the central order of Fig. \ref{fig:orders}  }    
   \label{fig:centralorder}
 \end{figure}

In Fig. \ref{fig:orders} we show, in arbitrary units, the value of the intensity recorded in the central pixel of the camera during the complete spectral scan of 4096 steps. Over a complete spectral scan the distance between the plates of the ET50 increases about 
1.9 $\mu$m, and 
every CCD pixel observes six interference orders. Figure \ref{fig:centralorder} provides an enlarged view of the central interference order. 
%The same intensity plots are shown in \citet{2005PASP..117.1435D}
%Denker \& Tritschler 2005 
%(their Fig. 3; although in transmission rather than reflection). 
%However, they only use the distance between orders to derive the 
%calibration in spectral tuning (i.e. m\AA/step) while we employ the whole curve to explicitly account for the possibility of 
%additional deformations of the cavity during the scan.

In both Figs  \ref{fig:orders} and \ref{fig:centralorder} 
the red points represent the experimental data, while the solid black line represents the best fit of Eq. \ref{eq:IRfit}
(to perform the fit we used the routine $\it{mrqmin}$ of Numerical Recipes
%\LEt{ is this a proper noun? If not, please remove the caps.}, 
based on the Levenberg-Marquardt method).
It is important to remark that the fit includes all the interference orders at once, and not only the central one as in \citet{2005PASP..117.1435D}
%Denker \& Tritschler (2005) 
and \citet{2008A&A...481..897R}.
%Reardon \& Cavallini (2008), 

Also visible in the Figures, 
Eq. \ref{eq:IRfit} describes the experimental data very well; indeed, for every pixel $(x,y)$  the rms
value of the differences between the observed and fitted intensity, normalized to the value $A(x,y)$ (Eqs. \ref{eq:ITfit} and \ref{eq:IRfit}) is limited to 1\% -- 3\%, 
%What is rms inside and outside of peaks? will indicate what is photon/read noise. 
thus validating the assumption contained in Eq. \ref{eq:SviF}. In particular, we find that the step $\Delta_1$ 
defined in Eq. \ref{eq:Delta1} and derivable from $\alpha(x,y)$ via the equation:

\begin{equation}
\centering
\Delta_1 (x,y) = \lambda /(4\pi) \times \alpha (x,y)
\label{eq:Delta1b} % Eq 16
,\end{equation}

\noindent
changes sensibly within the (recorded) cavity, having values that vary between 0.4605 nm and 0.4623 nm. 
Although the variation is only of order 0.5\%, note that its effect is amplified by the factor $n$ (Eq. \ref{eq:difetto}).

\begin{figure*} [ht]
\centering
% \vspace{12cm}
\includegraphics[width=16cm]{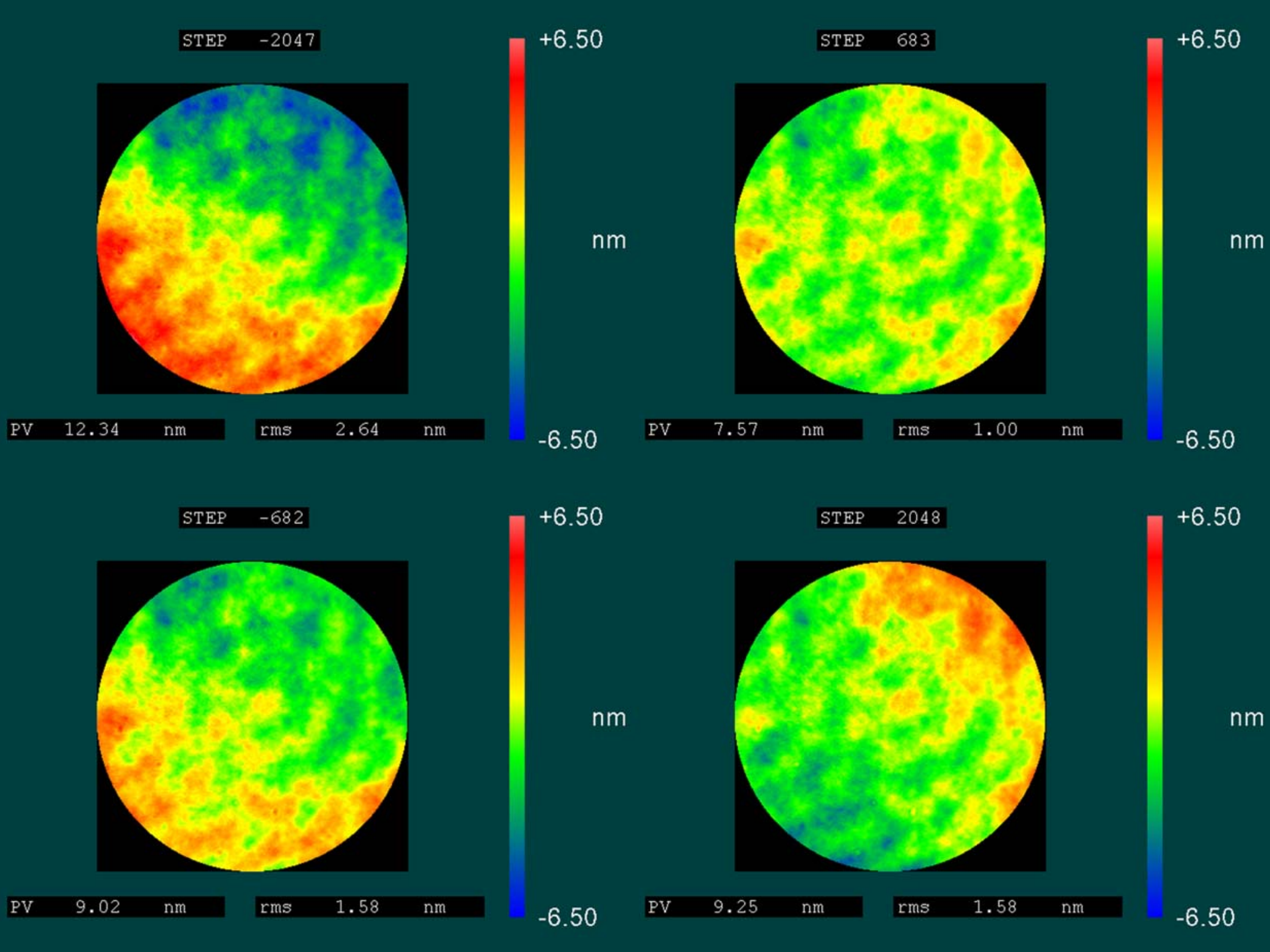}
   \caption{Maps of the cavity errors $M(x,y,n)$ for four values of the spectral step $n$, as indicated in the top of each map. A video of the cavity error maps for every spectral step is available in the online version of this paper (Fig6.mp4)}   
   \label{fig:4cavities}
 \end{figure*}

\subsection{Effects of spectral tuning} \label{sec:spectraltuning}

The fit of Figs. \ref{fig:orders} and \ref{fig:centralorder} was performed independently for the intensity curves acquired in each spatial
pixel $(x,y)$. %(each pixel independently of one another)
%with Eq. \ref{eq:IRfit} . 
Thus, after determining the six maps $A(x,y)$, $B(x,y)$, $C(x,y)$, $\alpha(x,y)$, 
$\beta(x,y),$ and $\gamma(x,y)$, we could calculate the map of the cavity errors $M(x,y,n)$ for every step of the scan 
(assuming H=$-$1; see Eq. \ref{eq:difetto}). Figure \ref{fig:4cavities} shows $M(x,y,n)$ for four different values of $n$ ($n$=$-$2047,$-$682,+683,+2048), while the corresponding
animation (available in the online version of the paper), displays the map for every spectral step.
 
 A large variation of the relative tilt of the two etalon plates at different spectral steps is visible in Fig. \ref{fig:4cavities}. This 
is surprising, since once the reference plane is defined (i.e., via the tilt optimization procedures described in the Introduction), the 
CS100 controller should keep the parallelism constant, by balancing the capacitance bridges that drive the 
piezo-electric actuators (Fig. \ref{fig:actuators}  below). Instead, we observe an obvious variation of the cavity during the scan, and in particular:

\begin{itemize}

\item The peak-to-valley $PV$ value, and the rms of the cavity errors vary sensibly, that is, within the intervals: 
\begin{equation} \label{eq:PVbeforelookuptable} %Eq 17
\centering
7.4~$nm$ ~\le PV \le12.3~$nm$    ;  ~~~~~1.0~$nm$ ~\le rms \le 2.6~$nm$
\end{equation}
\item There is an obvious tilt component, that varies both in amplitude and direction (see Fig. \ref{fig:tiltandangle} below);
\item There is a small-scale component that seems to remain constant during the scan. 
\end{itemize}

By fitting a plane to the cavity errors maps of Fig. \ref{fig:4cavities} we derived the amplitude and direction of the tilt as a 
function of scan step;  in Fig. 
\ref{fig:tiltandangle} 
we show how both the $PV$ and angle of the tilt vary as a function of $n$. The trend is extremely smooth for both quantities, with the local variations essentially contained within the width of the line in Fig. \ref{fig:tiltandangle}. 

The tilt $PV$  reaches a minimum ($\sim$ 1 nm)
around two thirds of the scan ($n$=693), while growing almost linearly toward the extremes of the scan. 
It reaches the maximum value of 9.8 nm at $n$=$-$2047. The tilt angle instead remains almost constant for half of the scan, 
and then grows  rapidly with a change of about 160$^\circ$. 
The two curves taken together seem to imply a continuous over- (or under-) correction by the piezo-actuators in a given direction, with the rapid variation of the tilt angle by almost 180$^\circ$ signifying the pivoting of the tilt plane around the minimum position. 
Both the smoothness of the effect, and its repeatability (see Sect. \ref{sec:correction} below)  appear consistent with an 
elastic deformation of the cavity. %due to the uneven action of the actuators. 

\begin{figure} 
\centering
\includegraphics[width=9cm]{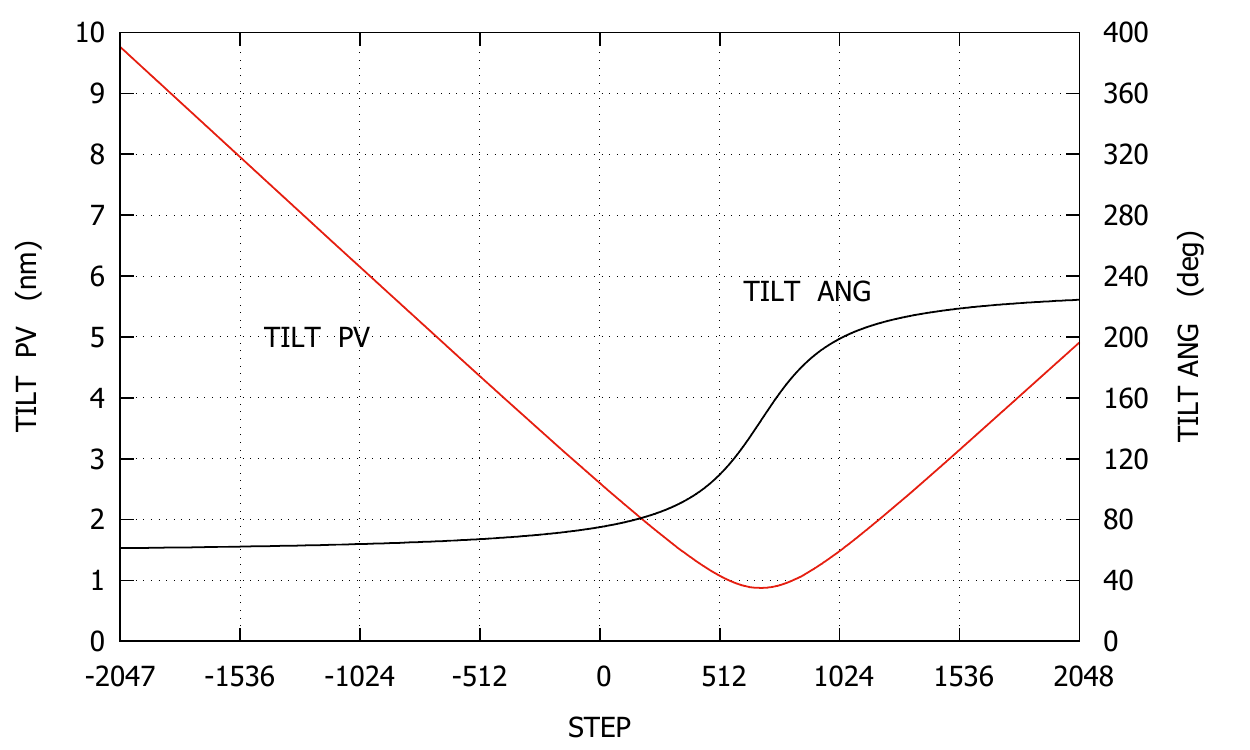}
   \caption{$PV$ and angle of the tilt component of the cavities of Fig. \ref{fig:4cavities} (TILT PV and TILT ANG, respectively), along 
   the full spectral scan. The curves represent the actual values of the parameters derived from the planar fit; 
   both vary very smoothly with $n$. Any local variation of the parameters is contained within the thickness of the font.}    
   \label{fig:tiltandangle}
 \end{figure}

As stated above,  before the spectral scan we took care to minimize the tilt of the cavity by changing 
the CS100 parameters, visually controlling the map of cavity errors for the distance corresponding to the central step $n$=0. As 
shown in Fig. \ref{fig:tiltandangle}, this corresponds to a tilt with $PV \approx$ 2.5 nm, which is probably around the best
precision that can be obtained with a visual procedure (cf. discussion in the Introduction). The spectral scan still introduces a significant, additional
component that much exceeds the starting value; using the simplified formula of Eq. \ref{eq:DeltaFWHM} and the values of Table 1, 
the extremes of the tilt curve of Fig. \ref{fig:tiltandangle} would correspond to broadenings of the spectral profiles for this FPI in 
excess of 
100\%. This would pose a problem for observing programs that require a large fraction of the full tuning range, for 
example for the case of broad solar chromospheric lines that are often sampled over several Angstroms (which corresponds to the total wavelength span of a cavity of a few mm gap, in the visible range). 
In both the classical and telecentric mount, this would, in different ways, result in variations in the instrument profile, creating an increasingly broader and more asymmetric instrumental profile as the scan progresses, away from the optimal, tilt-minimum position. It might also cause the spatial PSF to further vary as a function of wavelength tuning position. 
%The resulting instrumental profile towards the edges of the scan would be significantly broader with respect to the one pertaining to the middle of the scan,  although this effect might be masqueraded by the very shallow intensity gradients typical of the far wings of chromospheric lines. 

At the same time, we note that the case of Figs. \ref{fig:4cavities} and \ref{fig:tiltandangle} might represent
an extreme situation. As shown in Appendix, we measured the cavity defects of two more ICOS  ET50 (on loan from the National Solar Observatory), and in
both cases we derive an increase of the tilt when moving to the extremes of the spectral scan. However, the 
amplitude of the maximum tilt was sensibly smaller than for our case study: for these ET50s
%, of more recent fabrication, 
we obtained a tilt $PV$ = 0.5 nm for $n$=0, and $PV \approx$ 3 nm at the edge of the scan.
Nevertheless, the occurrence of such an effect in multiple similar devices points toward
an intrinsic property of these instruments, although we have no direct knowledge of the underlying cause. Since the capacitance 
bridges should drive the actuators until balanced, it seems plausible that the plates themselves (to which the capacitors are 
attached) might undergo elastic deformations during the scan.
%possibly due to a differing dynamical behavior of the piezo-electric actuators. 

\subsection{Stationary cavity defects} \label{sec:stationary}

\begin{figure} 
\centering
\includegraphics[width=8cm]{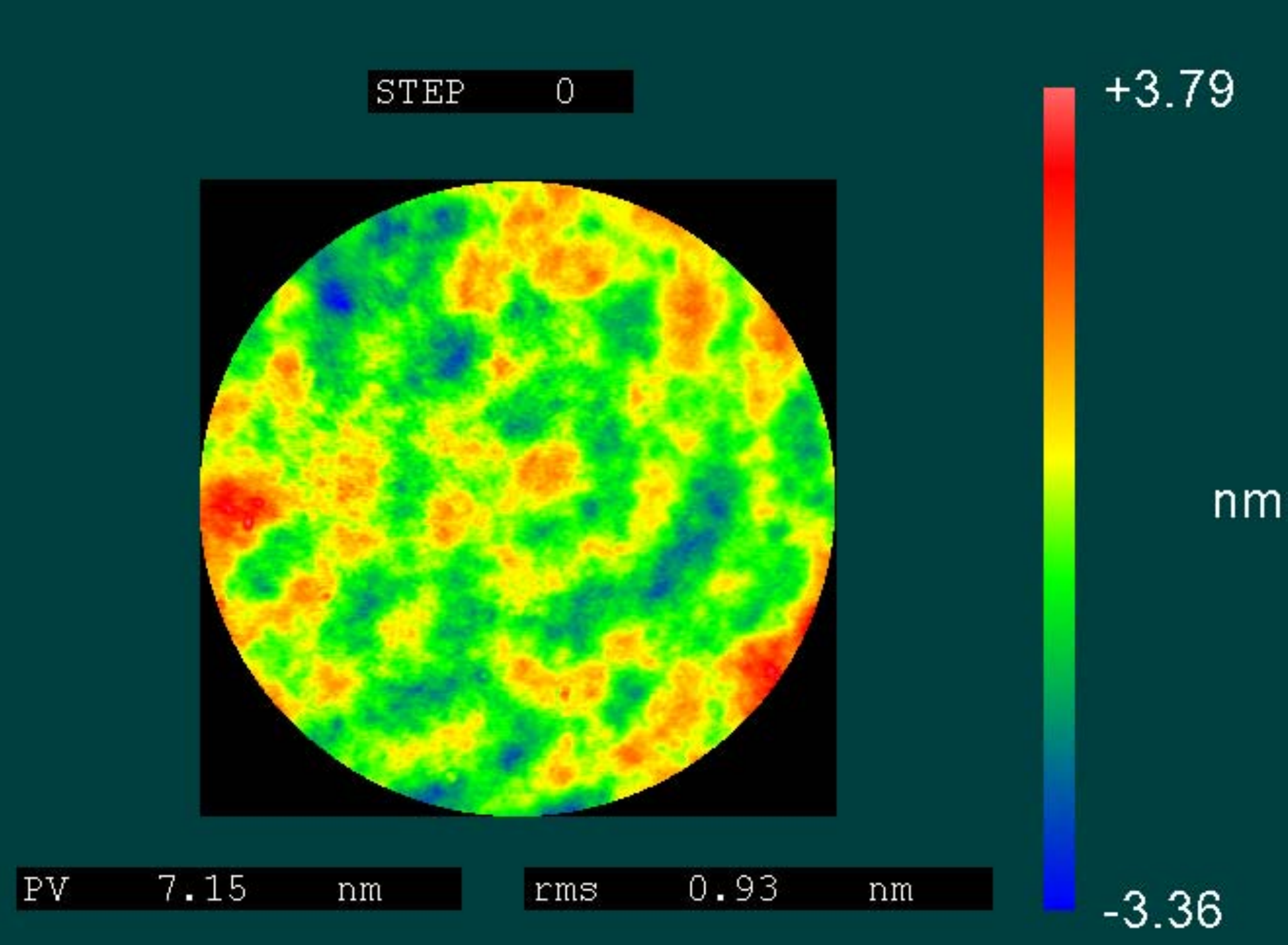}
   \caption{Map $M(x,y,0)$ for the central step of the spectral scan, after removal of the global tilt.}    
   \label{fig:stationary}
 \end{figure}

As mentioned above, if we remove the large-scale tilt from the cavity errors map $M(x,y,n)$), we find that 
the residual shape remains essentially constant throughout the scan, with a residual variation of less than 0.1 nm ($PV$). 
We can then use any arbitrary scan step to analyze the behavior (and possible cause) of the stationary cavity defects. 
Figure \ref{fig:stationary} shows the map $M(x,y,0)$ for the central $n$=0 step, after subtraction of the tilt plane. 

\begin{figure} 
\centering
\includegraphics[width=8cm]{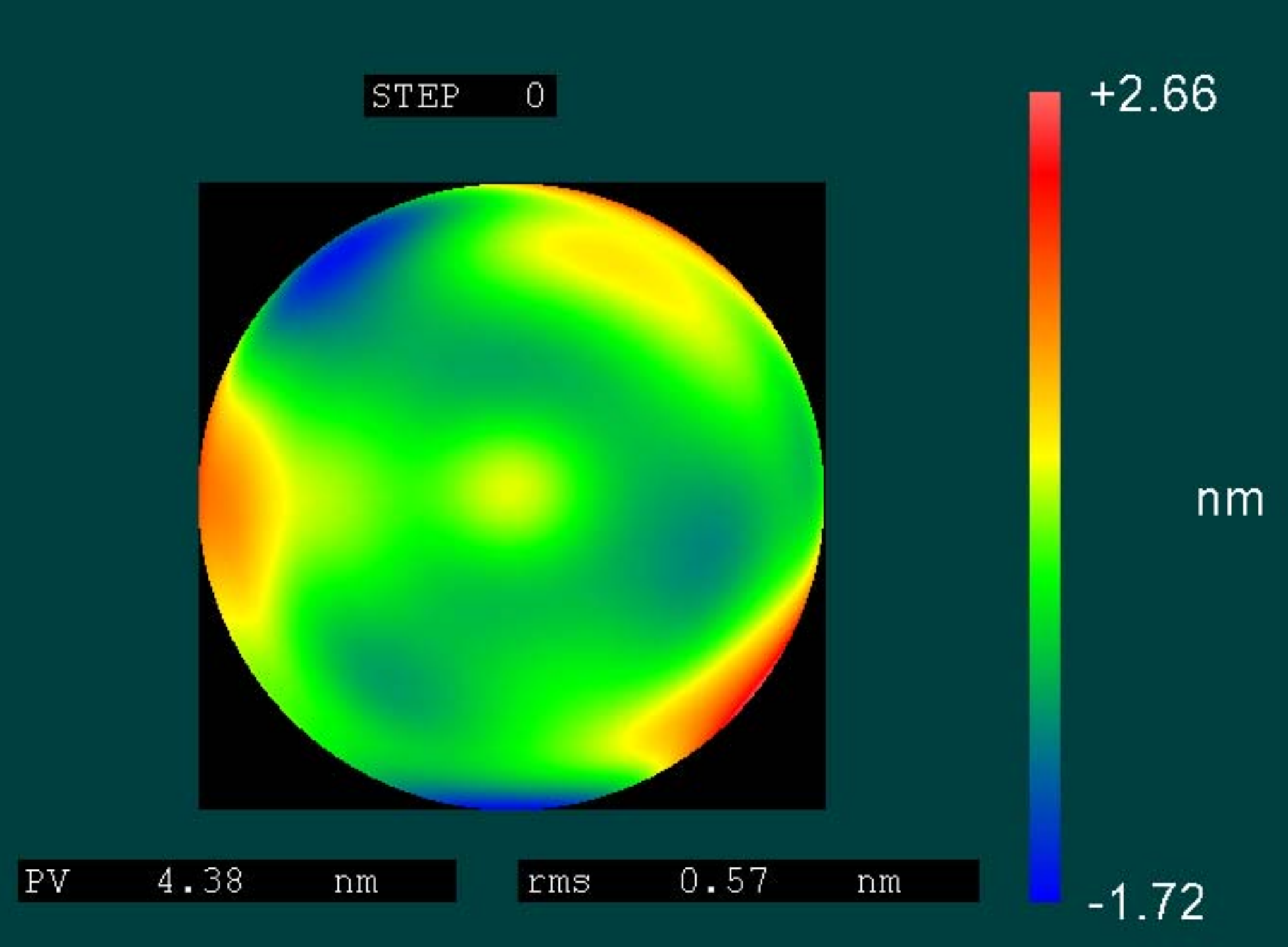}
   \caption{Fit of map $M(x,y,0)$ with the standard 37 Zernike polynomials of the FRINGE subset.}    
   \label{fig:Zernike}
 \end{figure}

\begin{figure} 
\centering
\includegraphics[width=8cm]{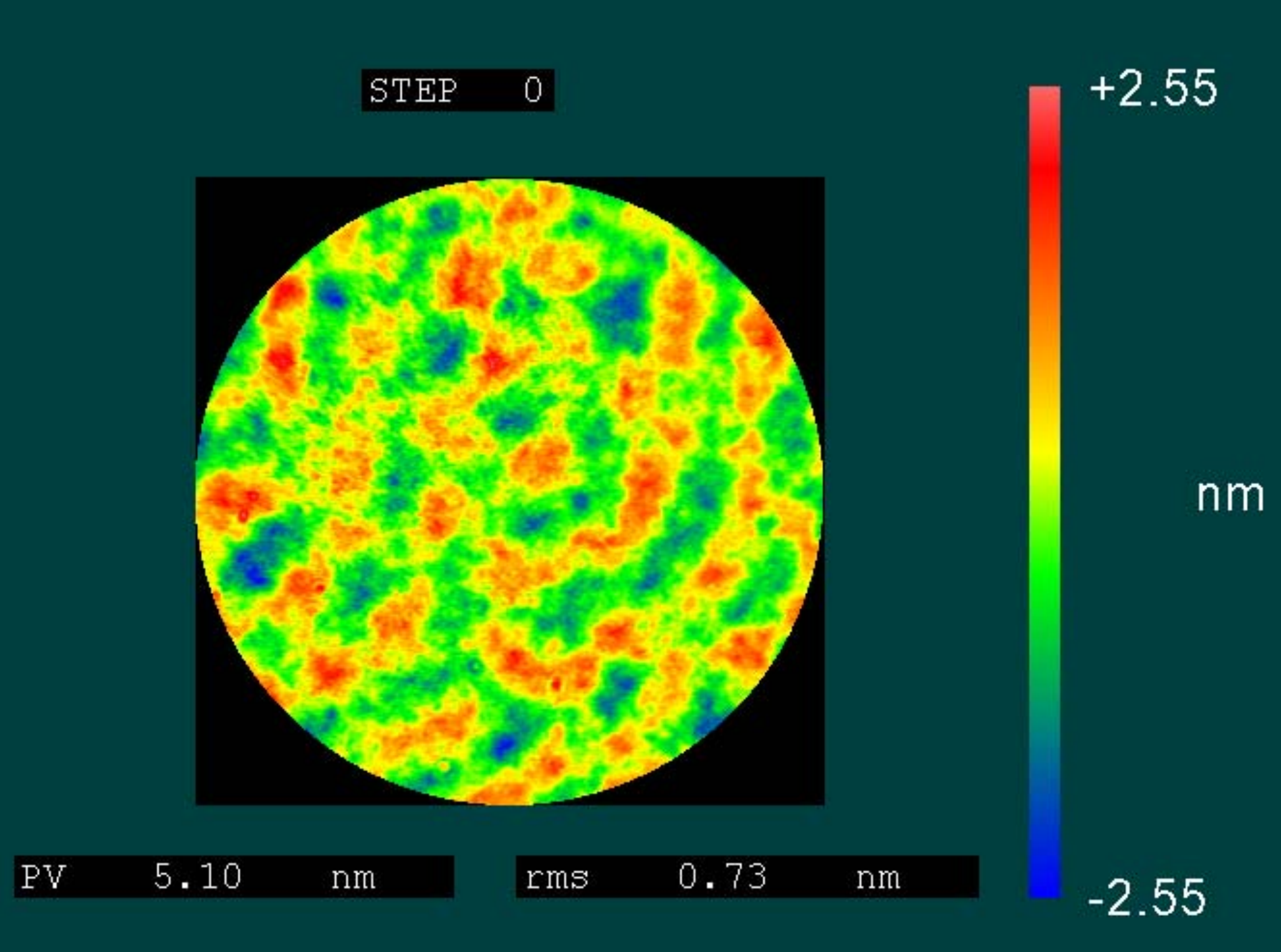}
   \caption{Residuals of M(x,y,0) after removal of tilt and the fit of Fig. \ref{fig:Zernike}.}    
   \label{fig:residuals}
 \end{figure}

We separated the defects at large scale (low spatial frequency) from those at small scale (high frequency) 
fitting the map of Fig. \ref{fig:stationary}  with the standard 37 Zernike polynomials of the FRINGE subset 
\citep{1992aooe...11....2W}.
%(J. C. Wyant and K. Creath, 1992). 
The resulting fits and residuals are shown in Figs. \ref{fig:Zernike} and \ref{fig:residuals} respectively.

The large scale errors (Fig. \ref{fig:Zernike}) describe a cavity with a roughly circular symmetry at the scale
of the plates, with peak-to-valley values of 
4--5 nm. This appears broadly consistent with the values measured by \citet[][their Figure  10]{2008A&A...481..897R}
%Reardon \& Cavallini (2008; ) 
over the central 35--40 mm of the 
IBIS' FPIs,  and with the ``bowing'' effect often observed in systems of this kind, due to the strong surface tension of the coating. The effect, however, is 
%measurably\LEt{ please check .} 
smaller than those introduced by the residual tilt at the extremes of the spectral scan.
%The amplitude of the effect is however not , although the effect?. 
%%GC metti qualcosa sulla preparazione delle lamine PRIMA del coating (effetto contrario)

We hypothesize that the three yellow and 
%or\LEt{ and? and/or? the slash is ambiguous.} 
red features (larger cavity gaps) positioned  around the perimeter at roughly 120$^\circ$
intervals are due to the local stress imparted by the piezo-electric actuators, that are equally spaced around the plates (see 
Fig. \ref{fig:actuators} below). 
A very similar tri-lobate pattern is measured for the large scale, static cavity errors of one of the NSO FPIs described in Appendix, while for the other we measure a ring-like figure. The bow-shape mentioned above is prominent in one of 
these two additional FPIs, reaching 2 nm $PV$ (comparable with the tilt measured at the extremes of the spectral scan), while almost negligible in the other.
Given that the FPIs are fabricated by the same manufacturer following similar procedures, this highlights how many different factors can determine the final cavity shape, including pre-load stresses
applied during assembly of the etalons, and the coating process.
% might introduce stresses that would be identified in the large scale error map; different coatings could then produce substantially different cavity shapes, whether or not the piezo give an imprint
Indeed, the cavity shape presented in Fig. 8
of \citet{2005PASP..117.1435D},
%Denker \& Tritschler 2005 
also obtained by applying the standard Zernike polynomials fit, is  different still. However, these authors present a fit to the global cavity, before any large scale defect (such as a tilt) is removed; this can effectively mask subtler effetcts
like those displayed in Fig.  \ref{fig:Zernike}. 

The small scale defects (Fig. \ref{fig:residuals}) range within $\pm$ 2.5 nm, again consistently with the results of \citet{2008A&A...481..897R}.
%Reardon \& Cavallini (2008). 
They show a uniform distribution throughout the cavity, with alternating positive and negative regions of average diameter of about 3 mm, probably due to
the process of coating deposition (see comments at the beginning of Sect. \ref{sec:casestudy}).

\section{Dynamic correction of the tilt} \label{sec:correction}

As described above, the ``spurious'' planar tilt in the cavity shape (Fig. \ref{fig:tiltandangle}) introduced by the spectral scan results in significant variations of the instrumental profile. Thus, we attempted to devise a software procedure to minimize the tilt component by actively controlling
the plates' parallelism during the spectral scan, using the ICOS CS100 controllers. To our knowledge, this is the first time that 
such a correction has been considered for use in an operational instrument based on FPIs. 
% CHECK THE VELLIUEX 2010 AND WILLIAMS 2016 papers (SPIE) - indeed it seems the first time

\begin{figure}
\centering
\includegraphics[width=8cm]{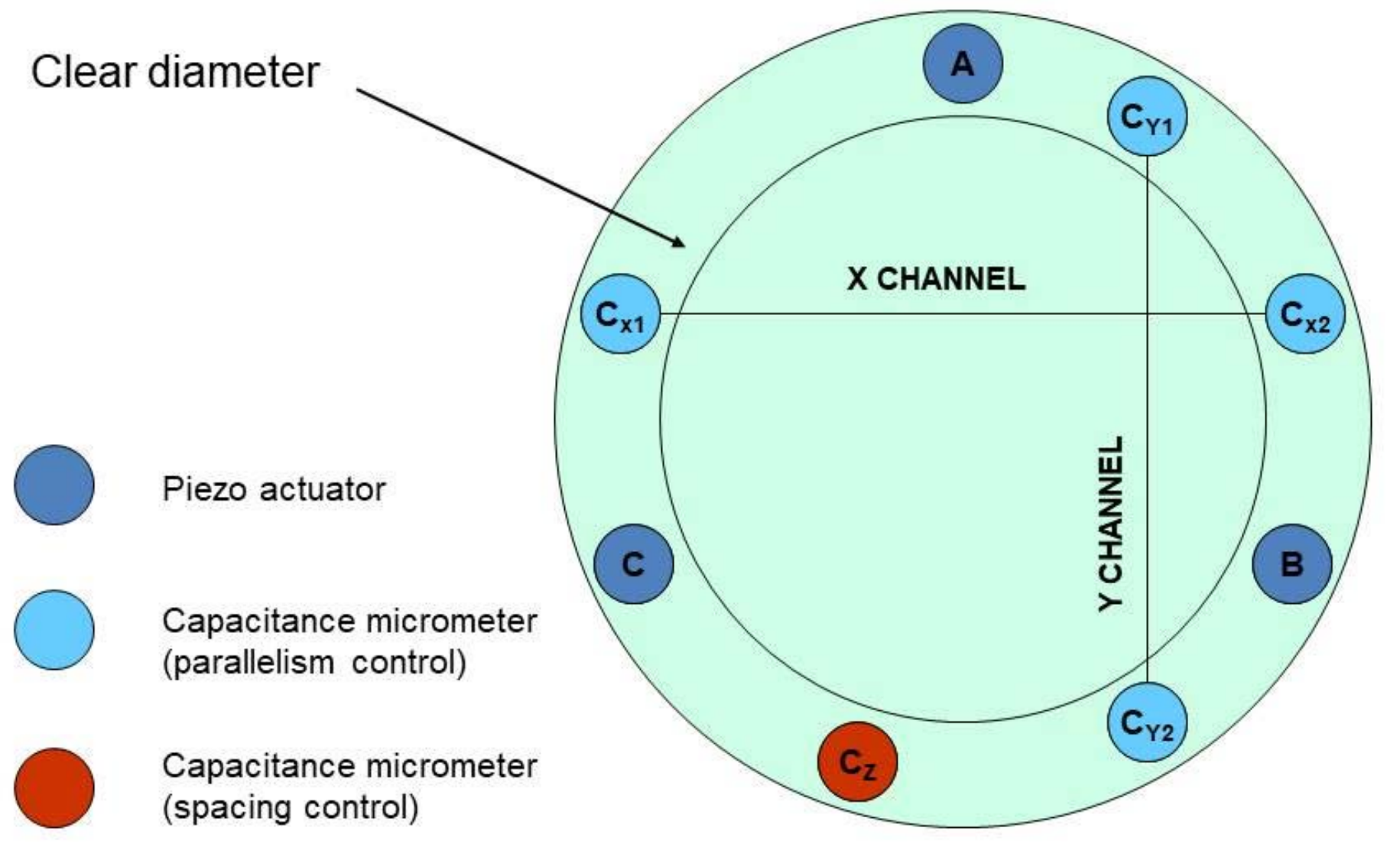}
 \caption{Scheme of the piezo-electric actuators and the capacitance controls for separation and 
 parallelism of the ET50' plates.}    
   \label{fig:actuators}
 \end{figure}

\begin{figure}
\centering
\includegraphics[width=8cm]{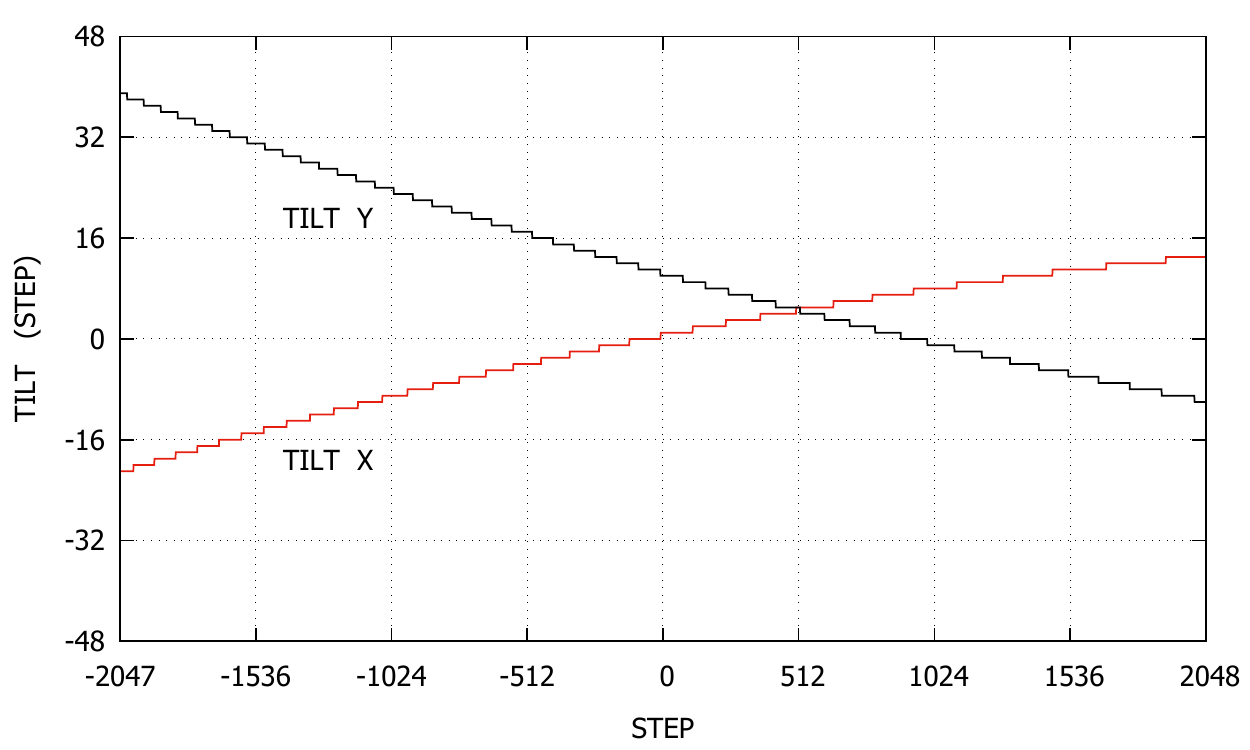}
   \caption{Lookup table of the (TILTX, TILTY) corrections  adopted to minimize the cavity tilt introduced by the scan.
 Each tilt step corresponds to  $1.4 \dot 10^{-3}$ arcsec. }
   \label{fig:lookuptable}
 \end{figure}

%GC Sfruttare in pieno la capacita' di CS100 di cambiare al volo tilt x e tilt y !!!!
Figure \ref{fig:actuators} represents schematically the actuators' positions, and the capacitance controllers for both 
separation and parallelism, for the case of the ICOS ET50 etalon. As discussed in Sect. \ref{sec:spectraltuning},
%3\LEt{ you may prefer to specify the subsection.}, 
once the reference 
plane is defined, the balancing action of the capacitance bridge should drive properly the piezo-electric actuators and maintain the parallelism for every value of the separation; however, unknown effects partially disrupt this feedback mechanism, with the net result of a varying tilt between the plates.  An important characteristics of the
CS100 controller however, is the possibility to provide independent values of the x and y tilt components [TILTX, TILTY]
 to the plates,  effectively allowing the user to reset the reference plane at any given spectral step \citep[See e.g.,][Sect. 3.2,
for further description of the CS100' operation]{2010AJ....139..145V}.
%Veilleux et al 2010). 
Exploiting this capability, and having measured
the actual cavities at each step as described in Sect. \ref{sec:spectraltuning}, we defined a lookup table that associates a pair of (TILTX,TILTY) values to
each step $n$ of the scan, so to minimize the overall tilt. The granularity of the (TILTX, TILTY) settings is the same as the step size $\Delta$ ($\sim$ 0.5 nm), allowing a precision in the reference plane of $\Delta_\theta \approx 1.4 \times 10^{-3}$ (Table 1).
%$Like $n$, also TILTX and TILTY are digitized at 12 bits ($-$2047, 2048), with  $\Delta_\theta \approx 1.4 \times 10^{-3}$ arcsec (Table 1). 
The resulting lookup table is given in Fig. \ref{fig:lookuptable}.

\begin{figure*} [h]
\centering
% \vspace{12cm}
\includegraphics[width=16cm]{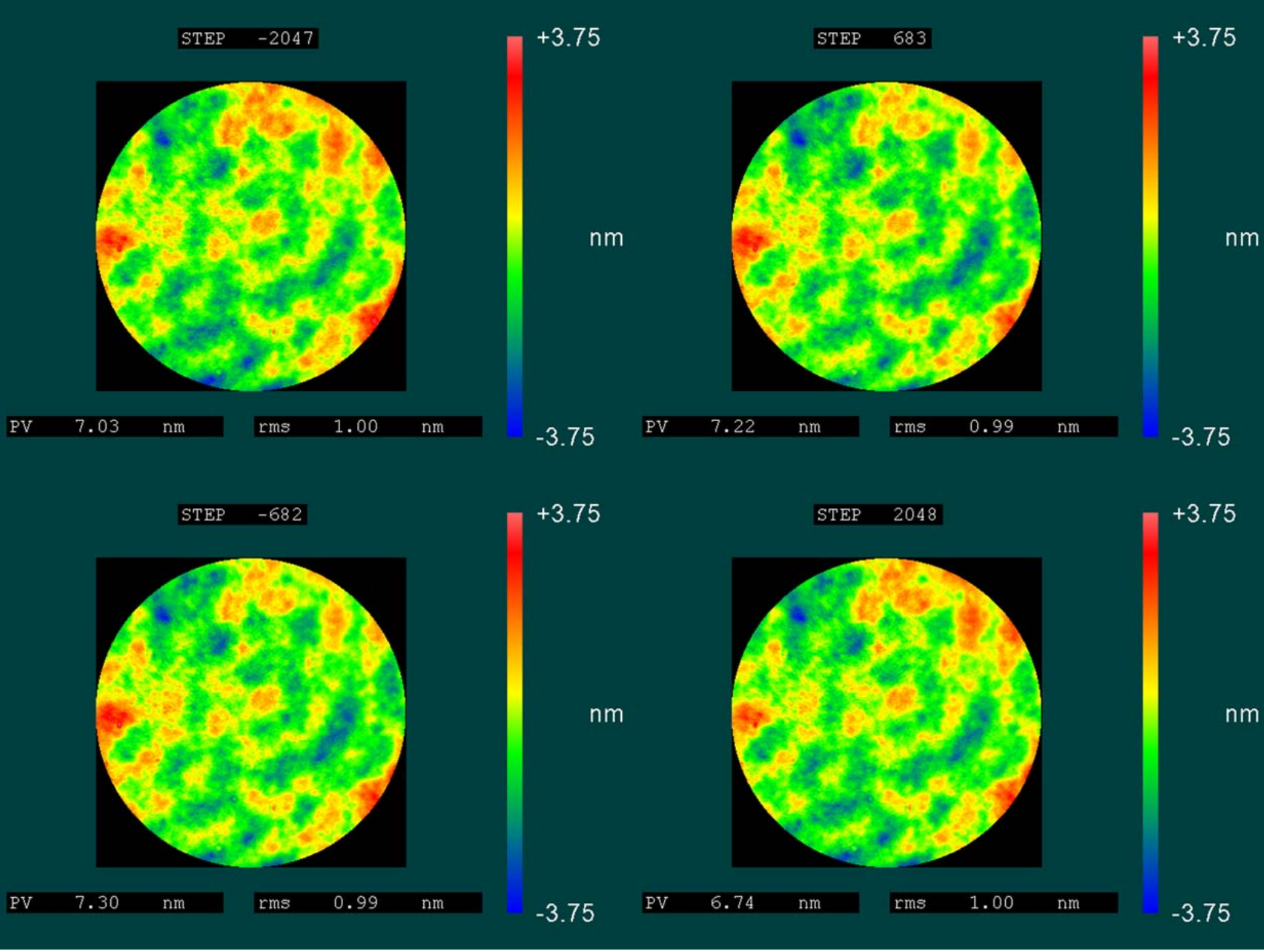}
   \caption{As Fig. \ref{fig:4cavities}, but after tilt minimization, as described in the text. The color scale is essentially identical to that of Fig. \ref{fig:stationary}. A movie of the cavity error maps  $M^\prime(x,y,n)$ for every spectral step is available in the online version of this paper (Fig13.mp4)}    
   \label{fig:4cavitiesafter}
 \end{figure*}

We then again acquired an interferogram at each spectral step, while applying the corrections defined by the lookup table. 
No extra delay is introduced in the scan, as the software controlling the FPI scan now simply transmits a string of three values: $n_z$, $n_x=$TILTX$(n_z)$, and $n_y=$TILTY$(n_z)$, instead of the single $n_z$ one. 
The
same fitting procedure described in Sect. \ref{sec:spectraltuning} was applied to the resulting intensity curves, and new maps $M(x,y,n)$ of cavity defects were obtained. Analogously to Fig. \ref{fig:4cavities}, Fig. \ref{fig:4cavitiesafter} shows the results for the same four different spectral steps
% ($n$ = $-$2047, $-$682, $+$683, $+$2048), 
after the automated tilt minimization procedure. The animation Fig13.mp4, available in the online version of the paper, shows the resulting map for every spectral step.

\begin{figure} 
\centering
%  \vspace{8cm}
\includegraphics[width=9cm]{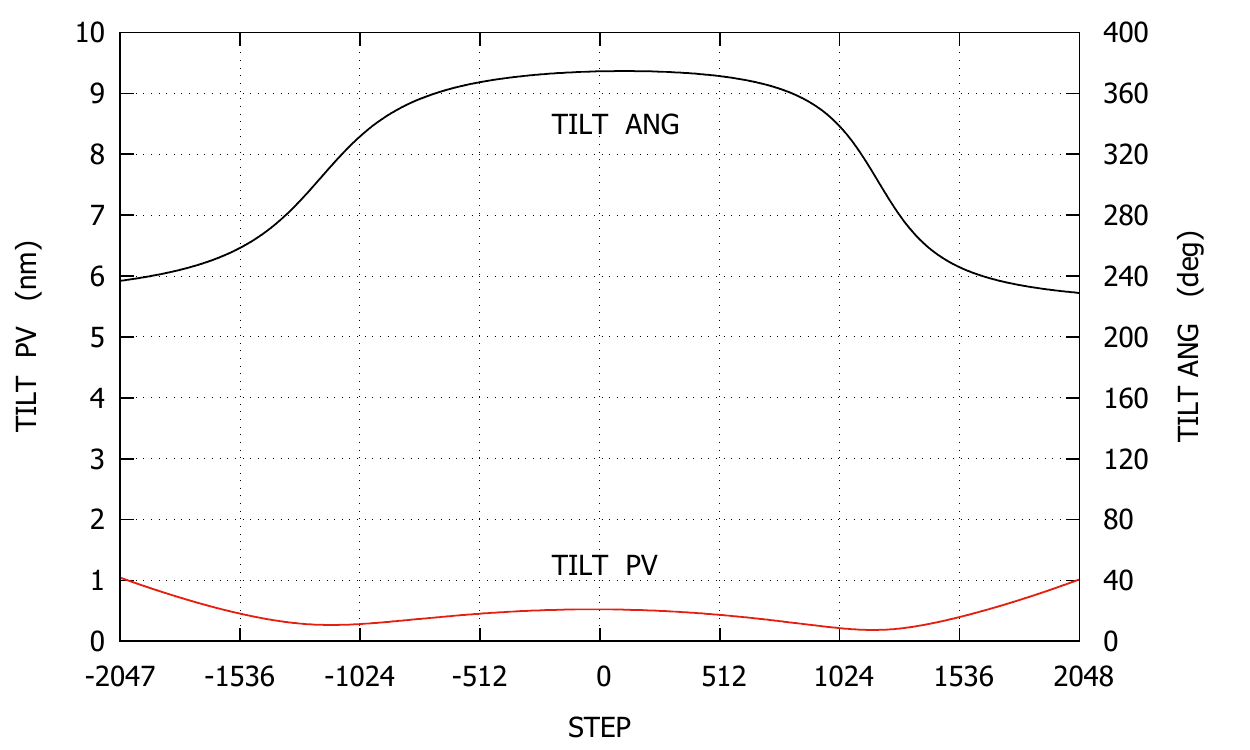}
 \caption{$PV$ and angle of the tilt component of the cavities of Fig. \ref{fig:4cavitiesafter} (TILT PV and TILT ANG, 
   respectively), along 
   the full spectral scan. The curves represent the actual values of the parameters derived from the planar fit; 
   both vary very smoothly with $n$. The tilt $PV$ remains within the 1 nm value.}    
   \label{fig:finaltilt}
 \end{figure}

The variation of the cavity shape is now strongly reduced; indeed, next we measured a $PV$ and  rms value of the cavity 
defects that vary within 

\begin{equation}\label{eq:PVafterlookuptable} %Eq 18
\centering
6.7~$nm$ ~\le PV \le 7.3~$nm$    ;  ~~~~~0.980~$nm$ ~\le rms \le 1.0~$nm$
,\end{equation}

\noindent
which are  sensibly smaller than the original values of Eq. \ref{eq:PVbeforelookuptable}.         
In particular, by fitting the new cavity 
defect maps as done in Sect. \ref{sec:spectraltuning} above, we find that the residual tilt after the correction procedure remains smaller
than 1 nm, and its direction does not increase monotonically. Rather, after rotating about 150$^\circ$ in the first half of the scan, 
the sense of rotation is flipped, and the direction returns essentially to the initial value (Fig. \ref{fig:finaltilt}). 

\begin{figure*} [h]
\centering
% \vspace{12cm}
\includegraphics[width=9cm]{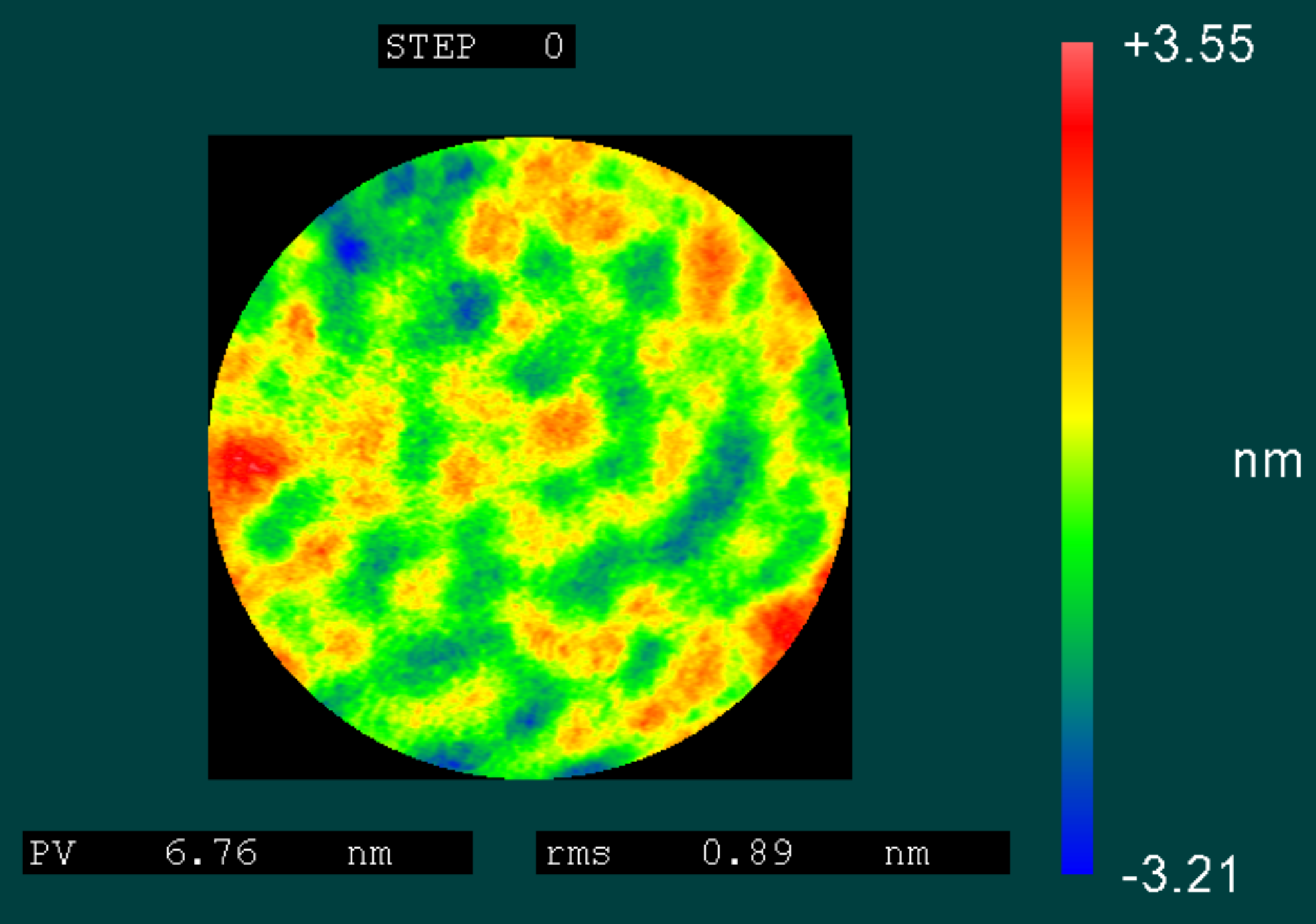}
   \caption{Cavity defects map $M(x,y,0)$ obtained applying the tilt minimization technique described in the text 
   %{\it \LEt{ please remove the italics.}
   four months
   after deriving the corrective lookup table. }    
   \label{fig:mapafter4months}
 \end{figure*}

In order to estimate how reliable and reproducible the correction technique is (and hence the stability of the system),
we repeated the same measures after an interval of four months, using the same initial settings of the CS100, and 
the same lookup table of Fig. \ref{fig:lookuptable}. Fig. \ref{fig:mapafter4months}
shows the resulting cavity defects map $M(x,y,0)$ at the central $n$=0 step. The derived map is essentially identical to that of Fig. \ref{fig:stationary}, testifying to the stability of the system and the reproducibility of the correction. The $PV$ value of the tilt (not shown) remains below 1 nm. 
A long-term stability of the parallelism of the plates has been noted earlier by \citet{2010AJ....139..145V},
%Veilleux 2010, 
that comment 
how the same tilt parameters could be applied from run to run over a period of years. 
This agrees with our experience in operating IBIS at the Dunn Solar Telescope over fifteen years.

\begin{figure*} [h]
\centering
% \vspace{12cm}
\includegraphics[width=16cm]{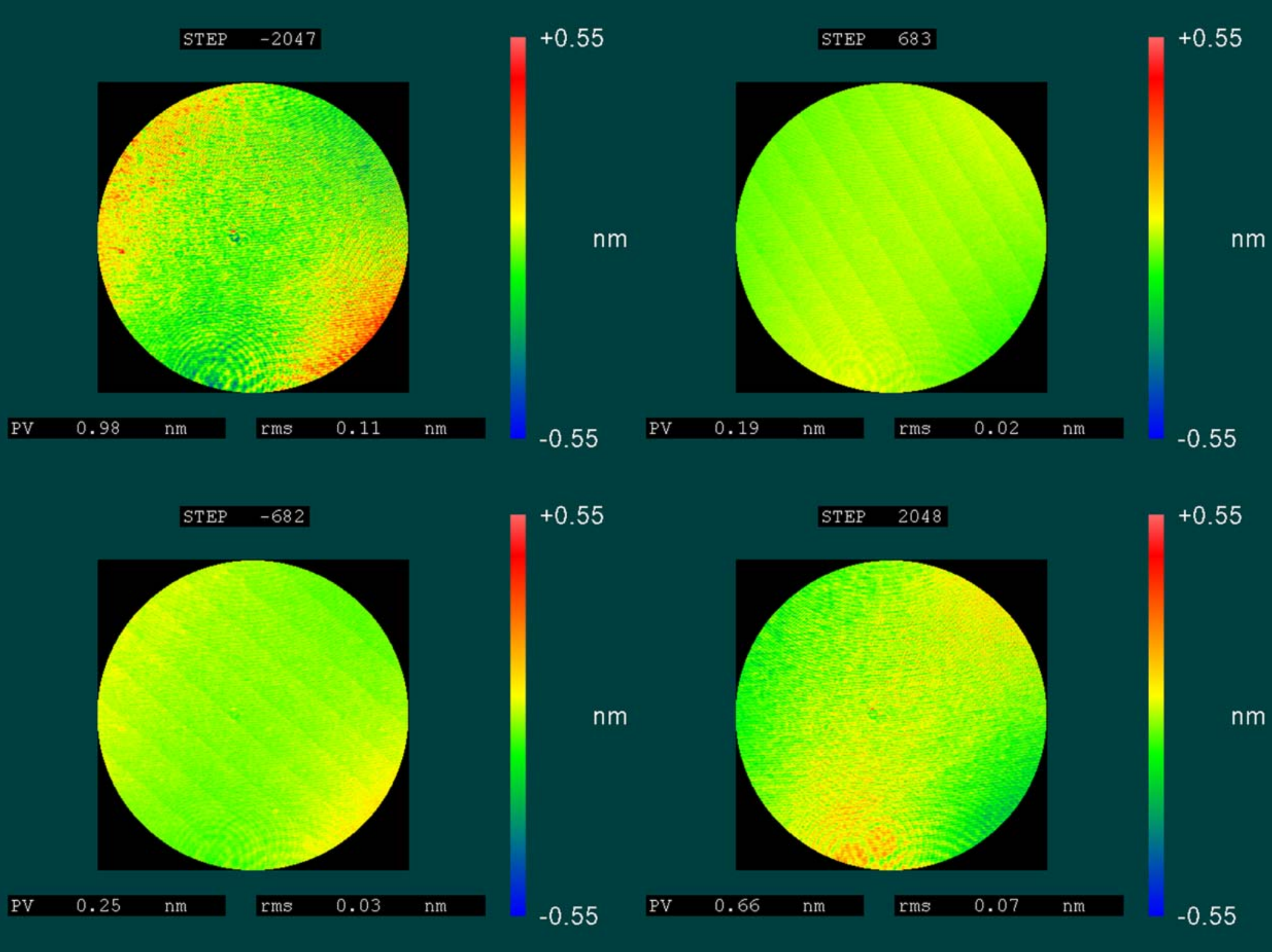}
   \caption{Cavity variations due to the deformation of the optical surfaces during a scan. The maps have been obtained by
   subtracting $M(x,y,0)$ from $M(x,y,n)$, after removal of the tilt. The small-scale fringes and straight patterns are due to
   residual interference between the optical surfaces, and from the CCD readout, respectively, and are not related to the shape of the cavity. }   
   \label{fig:astigmatism}
 \end{figure*}

As a final note, we also highlight how the technique described by Eqs. \ref{eq:ITfit} -- \ref{eq:difetto} is  sensitive enough to measure the cavity 
variations due to deformation of the optical surfaces during the scan. This was one of our original goals, as discussed in the Introduction. In Fig. \ref{fig:astigmatism} we show the difference maps 
$M(x,y,n) - M(x,y,0)$ for four different values of $n$ ($n$=$-$2047,$-$682,$+$683,$+$2048), after removal 
of the tilt component. The scale of variation is much smaller than that due to the tilt (about one order of magnitude), but 
still significant, and with some definite spatial structure. For example, it
is interesting to note how, during a scan, the piezo-electric actuators introduce a 
small astigmatism that rotates about 90$^\circ$ from one extreme to the other.  

\section{Discussion and conclusions} \label{sec:discussion}

%GC Attenta: bisogna rigirare la storia dei piezo. E' il capacitovo che in qualche maniera si modifica, non i piezo che risponsono
%male: questi sarebbero sovrascritti dal servo-loop
In this paper we have introduced a new technique to precisely measure the cavity defects of Fabry-Perot interferometers, which represent the core components of a large variety of astronomical instruments, both for nighttime and solar studies. The exact knowledge of the 
cavity
is a necessary step to properly understanding an instrument's performance (shape and variation of the instrumental profile) and interpret the observations. 

Our method, based on Eqs.  \ref{eq:ITfit}-\ref{eq:difetto},  is an extension of the phase-shifting 
techniques developed  to characterize optical surfaces  \citep{schreiber2007phase}
%Schreiber & Bruning 2007
and can be applied to FPIs with any value of the reflectivity R. The method can be used by measuring either the 
transmission or reflection interference pattern (Sect. \ref{sec:Zygo}); the latter is however more versatile, as it could be used also for the case of uncoated cavities, or
for wavelengths for which R is low. We envision, for example, using the same setup described in this paper (with a HeNe laser) to characterize cavities coated for near-infrared observations. 

The measures were performed at the Italian Istituto Nazionale di Ottica. Our 
setup, depicted in Fig. \ref{fig:figZygo}, was realized using a phase-shifting interferometer GPI-XP of Zygo-Ametek (www.zygo.com), that outputs a collimated HeNe ($\lambda$=632.8 nm) laser beam of 100 mm diameter. In a future work, we plan to 
employ an existing 150 mm laser beam expander to extend this study to FPIs of larger diameter.

Similarly to the 
work described by \citet{2005PASP..117.1435D} and \citet{2008A&A...481..897R},
%Denker \& Tritschler 2005, and Reardon \& Cavallini 2008, 
we used measurements of the reflected (transmitted) intensity from the HeNe laser to derive the cavity shape of an FPI mounted in a collimated configuration. As an extension of these works, however, we explicitly considered the possibility that the cavity shape can change during the spectral scan.
%, {\bf for example due to an imperfect} action of the actuators that actively control the spacing between the FPI plates, {\bf OR TO ???? }. 
We employed our technique to fully characterize the cavity of three separate ET50 interferometers (50 mm in diameter, coated for use in the visible range), all fabricated by ICOS Ltd.

Our most important result is that the cavity shape does indeed change significantly during a full spectral scan. Even after the ``standard'' 
alignment procedure, of the kind usually employed during instrumental setup at the telescope, a significant residual tilt is observed, 
that smoothly increases toward the edges of the scan. The effects of this residual tilt on the overall transmission profile can be 
severe: as shown in Sect. \ref{sec:spectraltuning}, in our worst example they can result in a doubling of the FWHM of the 
instrumental profile, depending on the position within the scan. We remark, however, that the effects appear substantially milder for the other two FPIs tested, which are of more recent fabrication and which, we hypothesize, might have a more uniform coating. Still, tests to characterize the cavity shape at 
%\LEt{ please remove the italics.}{\it 
each spectral step should be 
performed in order to properly characterize any given instrument.

To our knowledge this is the first time that this spectral dependence of the cavity shape is characterized. 
Interestingly, \citet{2010AJ....139..145V}
%Vielleux et al 2010 
reported a broad dependence of the tilt on the spacing for the case of the MMTF, but at a much 
coarser spectral resolution (i.e., $z$-axis settings of the CS100). Further, their results might also be affected by the dependence of the cavity shape on wavelength (see discussion below). 

The variation pattern of the measured tilt (Figs. \ref{fig:4cavities} and \ref{fig:tiltandangle}) appears consistent with an over- (or under-) correction of one of the actuators with respect to the other two, maybe because of an elastic deformation of the plates to which the capacitive sensors are attached. As the same qualitative behavior is observed for all the three ICOS ET50 tested, this points toward an intrinsic operational characteristics of these systems. Preliminary tests with a larger-format, uncoated interferometer also reveal a similar pattern, supporting our findings (Greco et al, in prep.). 
%{\bf With our data, it is not possible 
%to assess whether this effect is actually due to an imperfect action of the actuators themselves, or to a slightly imprecise
%measure of the cavity by the capacitance bridge. ANYTHING ELSE?}

To minimize the residual tilt during the scan, we devised a correction procedure that exploits the capabilities of the CS100 controller to dynamically input a value for the the tilts of both the $x$ and $y$ axes for each step of the spectral scan ($z$ axis). After applying the
measuring technique described above, and fitting a planar tilt to the cavity shape at each spectral step, we derive a lookup table
of the additional (TILTX,TILTY)  values necessary to minimize the tilt component for every step. The commands can be provided within the same 
controlling sequence used for the spectral scan, without introducing any additional delay. The results are highly 
encouraging, with the corrective sequence limiting the residual tilt to $PV$ values of less than 1 nm. Moreover, the system appears very stable: the same results were obtained
%replicated\LEt{ we replicated the same results?.} 
when applying the initial settings, and the lookup table described above, after a long period (four months). Together with the very slow dependence of the tilt on the step number $n$, this extreme repeatability points toward an elastic deformation of the cavity as mentioned above. The stability of ICOS FPI systems in terms of their 
setup parameters in operational situations has indeed been reported numerous times \citep[e.g.,][]
{2010AJ....139..145V};
%Vellieux 2010; ; 
this holds promise that our corrective procedure can be effectively adopted for existing and future instrumentation. 

Once the variable terms were taken into account, it was possible to uncover the ``static'' cavity shape, resulting from the overall fabrication process. This appears to be a combination
of two main components: a large-scale one, 
showing both a bow shape, typical of the strong surface tension of the coating and a tri-lobate figure most likely due to the local stresses introduced by the piezo
actuators, positioned at 120$^\circ$ from one another; and a small scale component (of size one to a few mm on the plates) probably due
to the coating deposition process. The overall deviations from cavity flatness are significant enough
to make it necessary to account for them in the calculation of the overall instrumental profile 
\citep{2008A&A...481..897R}. A different optical figure is however obtained for one of the three ET50 analyzed, probably due to differential
contributions of the many factors entering the fabrication and assembly of the etalons; this reinforces the need to obtain 
an exact measurement of each interferometer cavity to properly evaluate the instrumental transmission profile.  
%(Reardon \& Cavallini 2008). 

All of the results presented above have been obtained using a monochromatic HeNe laser at 632.8 nm, leaving open the 
question of whether they would hold at different wavelengths.  Indeed, several instances have been reported in the literature of operational FPI systems for nighttime astronomy
where the tilt optimal values were strongly wavelength dependent \citep{2008AJ....135.1825R,2010AJ....139..145V,2014MNRAS.443.3289G}. However, as already discussed by \citet{1998PASP..110.1059J} in their analysis of the TTF, this effect is mainly observed for etalons with
very small gaps (few to tens of $\mu$m), which are comparable to the thickness of the optical coatings.

Wavelength-dependent phase changes and non-uniformities should be less important at the large gaps usually employed in 
instruments for solar physics. It is, however, worth noting that a recent  study of the coatings of VTF \citep{2018SPIE10706E..1RP}
%(Pinard et al. 2018) 
mentions the possibility that phase differences at different laser wavelengths might exist. We plan to study these effects
in a future work by implementing the same optical setup, but with different laser sources. If tilt variations are a manifestation of true changes in physical plate separation, 
%{\bf due to variable response of piezoelectrics and capacitance stacks,} 
then we expect that this effect will manifest itself, additively, together with the other wavelength dependent sources of effective plate separation.\\

\begin{acknowledgements}
Part of this work was performed within the SOLARNET I3 integrated activity (European Union 7th Framework Program, 
grant No. 312495).  NSO is operated by the
Association of Universities for Research in Astronomy,  Inc. (AURA), under
cooperative agreement with the National Science Foundation.
We are grateful to Dr. V. Martinez Pillet, for lending us the two NSO ET50s used in additional testing, and to D. Gilliam for taking
care of all the practical issues related to their transport.
Finally, we would like to thank Chris Pietraszewski (IC Optical Systems, Ltd.) for invaluable discussions.
\end{acknowledgements}

\bibliographystyle{aa} % style aa.bst
%\bibliography{Greco.FPI} % your references Yourfile.bib

\begin{appendix} \label{sec:append}
\section{The cavity errors of additional ET50 interferometers}

To test our procedure on different etalons, we used two additional ICOS ET50 interferometers obtained on loan from 
NSO. The etalons have been fabricated in the mid 1990s, with cavity spacings of
2.8 mm and 609 $\mu$m respectively. Other relevant instrumental  
characteristics are comparable to those reported in Table \ref{tab:et50properties}; in particular the operational wavelength range is 400--700 nm as well, with a coating reflectivity of about 95\%.
%609 micron ET50FS 1001
%2.8 mm ET50FS 1046
% 400 -700 nm; 95%

\begin{figure*} [ht]
\centering
% \vspace{12cm}
\includegraphics[width=15cm]{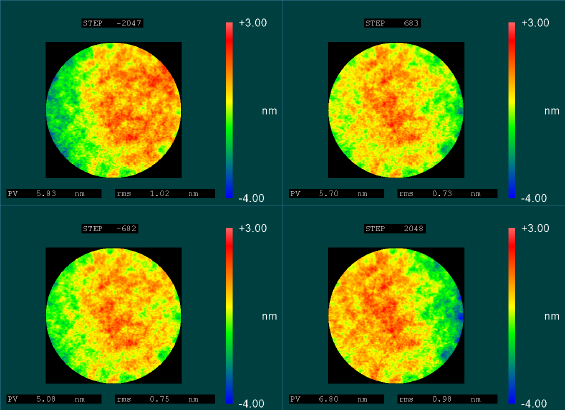}
   \caption{As Fig. \ref{fig:4cavities}, for the 609 $\mu$m gap NSO ET50.}   
   \label{fig:4cavitiesNSO1}
 \end{figure*}

\begin{figure*} [ht]
\centering
% \vspace{12cm}
\includegraphics[width=15cm]{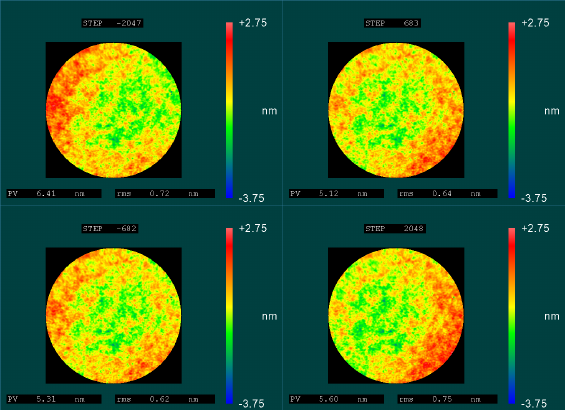}
   \caption{As Fig. \ref{fig:4cavities}, for the 2.8 mm gap NSO ET50.}   
   \label{fig:4cavitiesNSO2}
 \end{figure*}

For both ET50, we measured the reflected interferograms over a full spectral scan, as described in Section \ref{sec:casestudy}. 
The resulting images were analyzed by fitting the fringe pattern in each pixel's intensity curve  with Eq. \ref{eq:IRfit}. The 
resulting
cavity errors at four different spectral steps are represented in Figs. \ref{fig:4cavitiesNSO1} and \ref{fig:4cavitiesNSO2} for the
two FPIs, in the same way shown in Fig. \ref{fig:4cavities} for the interferometer studied in the main body of the paper.

\begin{figure} 
\centering
%  \vspace{8cm}
\includegraphics[width=9cm]{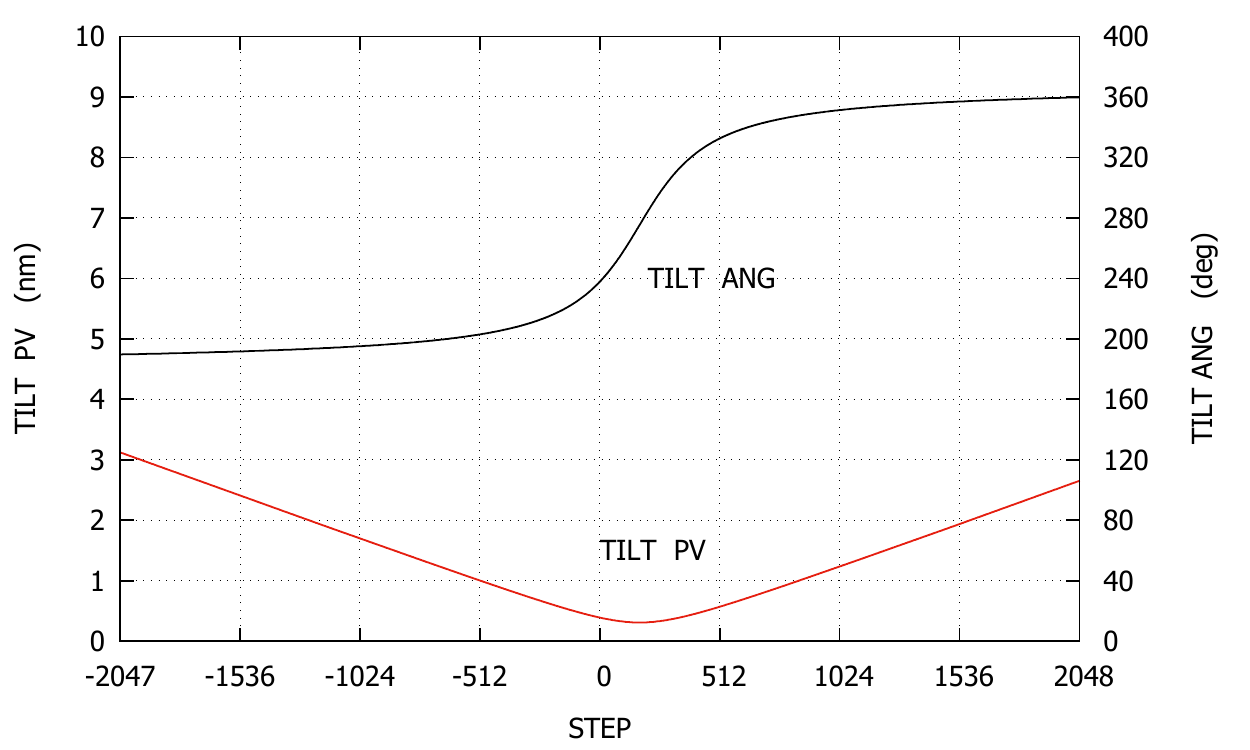}
   \caption{$PV$ and angle of the tilt component of the cavity error maps of the 609 $\mu$m gap NSO ET50 (Fig. 
   \ref{fig:4cavitiesNSO1}).}
     \label{fig:tiltandangleNSO1}
 \end{figure}

\begin{figure} 
\centering
%  \vspace{8cm}
\includegraphics[width=9cm]{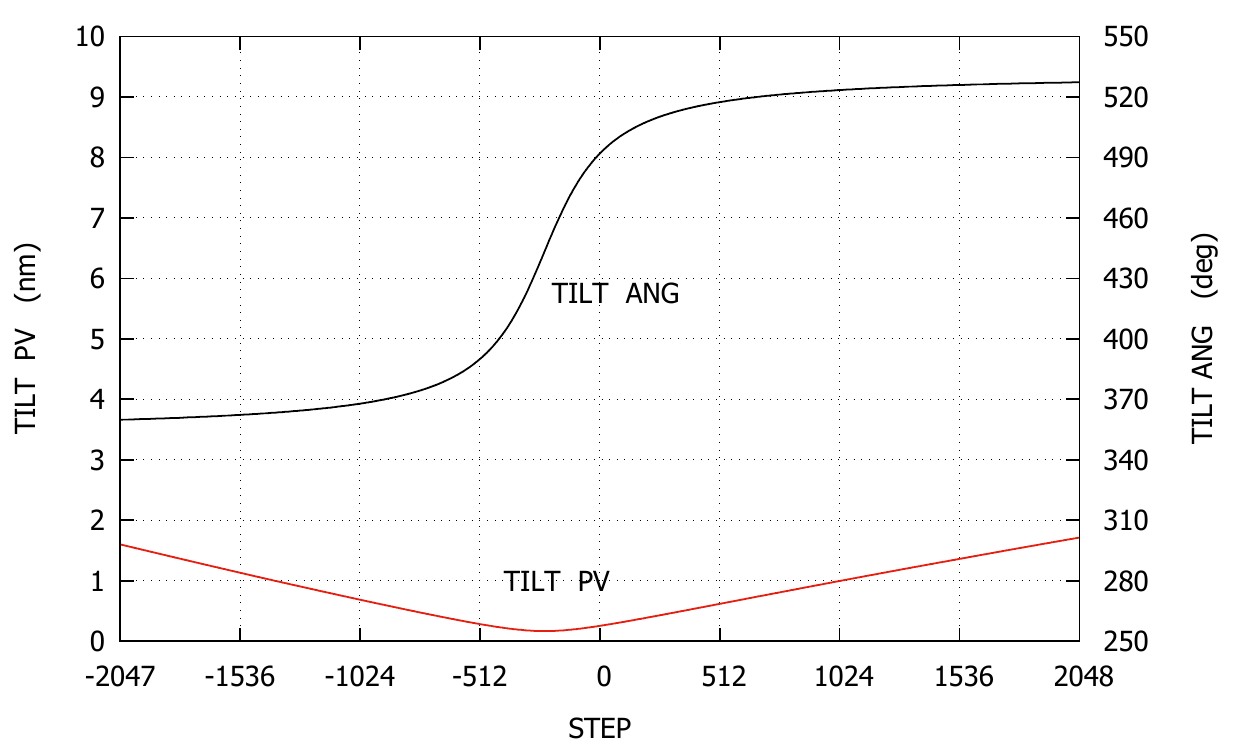}
   \caption{As Fig. \ref{fig:tiltandangleNSO1} for the 2.8 mm gap ET50  of NSO (Fig. 
   \ref{fig:4cavitiesNSO2}).}
     \label{fig:tiltandangleNSO2}
 \end{figure}

The results are consistent with what we report
%is reported\LEt{ or "what we report".} 
in Sect. \ref{sec:spectraltuning}. First of all, the $PV$ and $rms$ 
of the cavity errors vary sensibly within the spectral scan, with $PV$ values between 5 and 6.8 nm, and $rms$ values
between 0.7 and 1 nm for the first
NSO etalon; and $PV$ between 5.1 and 6.4 nm, and $rms$ between 0.62 and 0.75 nm for the second one. Further, an obvious tilt 
component is
visible in the cavity errors maps of both Figs. \ref{fig:4cavitiesNSO1} and \ref{fig:4cavitiesNSO2}; its evolution within the spectral scan essentially  causes the variation of the $PV$ values described above. As described in Sec. \ref{sec:spectraltuning}, we then
fit a plane to the error maps derived at each spectral step; the resulting parameters of the plane (tilt $PV$ value and angle) are
displayed in Figs. \ref{fig:tiltandangleNSO1} and \ref{fig:tiltandangleNSO2}.

The curves are qualitatively similar to those shown in Fig. \ref{fig:tiltandangle}, meaning that the tilt reaches a minimum value around the middle of the spectral scan (where initial parallelism is set), and increases monotonically toward the edges. The tilt angle flips 
sign around the minimum of the tilt $PV$ value, consistently with the idea expressed above of an over- or under- correction 
by the piezo-actuators in a given direction. The amplitude of the changes is however much smaller than for the IPM etalon.
%; this points towards a more accurate response of the actuators with the driving voltage. 
Still, as we describe in the Introduction,
with the typical values of reflectivity and working wavelength for these instruments, a residual $PV$ tilt value around 3 nm 
corresponds to a substantial broadening of the transmission profile. 

Here we do not show the subsequent elaborations (removal of the tilt, analysis of the static defects etc.) but they are fully
consistent with what shown for the IPM etalon in the body of the paper. Worth noting is the different spatial size of the residual
static defects, appreciable already in Figs. \ref{fig:4cavitiesNSO1} and \ref{fig:4cavitiesNSO2}. These are most likely due to 
variations in the coating substrate among the different interferometers and coating runs. 
\end{appendix}

\end{document}